\documentclass[%
reprint,
superscriptaddress,
 amsmath,amssymb,
 aps,
floatfix,
]{revtex4-1}
\usepackage{graphicx}
\usepackage{dcolumn}
\usepackage{bm}

\usepackage[flushleft]{threeparttable} 

\usepackage{dsfont}
\usepackage{color,psfrag}
\usepackage{mathtools,amscd}

\usepackage{soul}
\usepackage{xcolor} 

\usepackage{ifpdf}
    \ifpdf
\usepackage{tikz}
\usetikzlibrary{arrows,chains,matrix,positioning,scopes, fit, calc}
\usetikzlibrary{cd}
\usepackage{chemfig}
\fi
\usepackage{hyperref}

\hypersetup{
	colorlinks,
	linkcolor=blue,
	citecolor=blue, 
	filecolor=black,
	urlcolor=blue}

\newcommand{\mat}[1]{\mathbf{#1}} 

\newcommand{\tran}{^{\mathstrut\scriptscriptstyle\top}}

\begin{document}

\preprint{APS/123-QED}


\title{Molecular Force Fields with Gradient-Domain Machine Learning: Construction and Application to Dynamics of Small Molecules with Coupled Cluster Forces}
\author{Huziel E. Sauceda}
 \affiliation{%
 Fritz-Haber-Institut der Max-Planck-Gesellschaft, 14195 Berlin, Germany
}%
\author{Stefan Chmiela}
 \affiliation{%
 Machine Learning Group, Technische Universit\"at Berlin, 10587 Berlin, Germany
}%
\author{Igor Poltavsky}
\affiliation{%
 Physics and Materials Science Research Unit, University of Luxembourg, L-1511 Luxembourg, Luxembourg
}%
\author{Klaus-Robert M\"uller}%
 \email{klaus-robert.mueller@tu-berlin.de}
\affiliation{%
 Machine Learning Group, Technische Universit\"at Berlin, 10587 Berlin, Germany
}
\affiliation{%
Department of Brain and Cognitive Engineering, Korea University, Anam-dong, Seongbuk-gu, Seoul 02841, Korea
}%
\affiliation{%
Max Planck Institute for Informatics, Stuhlsatzenhausweg, 66123 Saarbr\"ucken, Germany}
\author{Alexandre Tkatchenko}
 \email{alexandre.tkatchenko@uni.lu}
\affiliation{%
 Physics and Materials Science Research Unit, University of Luxembourg, L-1511 Luxembourg, Luxembourg
}%
\date{\today}

\date{\today}

\begin{abstract}
We present the construction of molecular force fields for small molecules (less than 25 atoms) using the recently developed symmetrized gradient-domain machine learning (sGDML) approach [Chmiela \textit{et al.}, Nat. Commun. {\bf 9}, 3887 (2018); Sci. Adv. {\bf 3}, e1603015 (2017)]. This approach is able to accurately reconstruct complex high-dimensional potential-energy surfaces from just a few 100s of molecular conformations extracted from \textit{ab initio} molecular dynamics trajectories. The data efficiency of the sGDML approach implies that atomic forces for these conformations can be computed with high-level wavefunction-based approaches, such as the ``gold standard" CCSD(T) method. We demonstrate that the flexible nature of the sGDML model recovers local and non-local electronic interactions (e.g. H-bonding, proton transfer, lone pairs, changes in hybridization states, steric repulsion and $n\to\pi^*$ interactions) without imposing any restriction on the nature of interatomic potentials. The analysis of sGDML molecular dynamics trajectories yields new qualitative insights into dynamics and spectroscopy of small molecules close to spectroscopic accuracy. 
\end{abstract}
\pacs{Valid PACS appear here}
\maketitle



\section{Introduction} \label{intro}
Molecular force fields (FFs) constitute one of the most important tools in chemistry, biology, and materials modelling due to their remarkable value in understanding systems that range from small molecules, e.g. ethanol with 9 atoms, up to large proteins, aiding in the exploration and the discovery of new materials and drugs.
Creating physically inspired and handcrafted interatomic potentials with parameters fitted to experimental data or quantum-mechanical calculations has been a common practice since the early works on molecular dynamics~\cite{Alder1959,Rahman1964,Verlet1967,Rahman1971}.
The complexity of creating reliable interatomic interaction models using prior physical knowledge led to the development of dedicated specialized FFs
for different material classes, including tight-binding potentials for semiconductors and metals~\cite{EAM1984}, Tersoff potential for covalent materials~\cite{Tersoff1988}, polarizable FFs~\cite{PolarFF2006}, the TIP\textit{n}P FFs for water~\cite{TIP4P1983,TIP5P2000}, and a wide variety
of biomolecular FFs such as AMBER, CHARMM, MMFF, and GROMOS that often lead to reliable results for folded protein structures under ambient conditions~\cite{AMBER1981,CHARMM1983,MMFF94,GROMOS2005}.
The wealth of available interatomic potentials illustrates the vast amount of fundamentally different material classes, exposing that treating different types of interactions (metallic bonding, covalent chemistry, hydrogen bonding, non-covalent interactions, etc.) in a unified and seamless fashion is a complex challenge for handcrafted mechanistic FFs.

To resolve some of these challenges, a number of recently developed machine learned (ML) FFs exploit the redundant information contained in datasets of \textit{ab initio} calculations or molecular dynamics trajectories to
reconstruct the underlying potential energy surface (PES) without imposing any particular handcrafted analytical form for the interatomic potential.
 
In particular, a vast amount of work has been done in molecular representations~\cite{Rupp2012,Bartok2010,Hansen2013,Bartok2015_GAP,Hansen2015,Rupp2015,Ceriotti2016,artrith2017efficient,Ceriotti2017,Glielmo2017,yao2017many,faber2017prediction,eickenberg2018solid,glielmo2018efficient,Grisafi2018,tang2018atomistic,pronobis2018many,FCHL2018}, neural networks architecture development~\cite{Behler2007,Behler2012,Behler2016,Gastegger2017,dtnn,SchNet2018,SchNetNIPS2017,ryczko2018convolutional,zhang2018deep}, data sampling~\cite{DeVita2015,Shapeev2017,dral2017structure,noe2018,noe2018b} and inference methods~\cite{Bartok2013,Montavon2013a,Ramprasad2015,Brockherde2017,huan2017universal,Tristan2018,lubbers2018hierarchical,kanamori2018exploring,hy2018predicting,Smith2017,Clementi2018,winter2019,ResponseFieldAnatole2018,gdml,sgdml}, as well as software development~\cite{sGDMLsoftware2019,yao2018tensormol,schnetpack2018} and explanation methods~\cite{innvestigate,meila2018}.

Based on rigorous statistical learning theory~\cite{SLT1,SLT2}, machine learning provides a powerful and general framework for constructing force fields, since ML approaches can reconstruct complex high-dimensional objects with arbitrary accuracy, provided that sufficient data samples (molecular energies and atomic forces) are used for training. Obviously, the computational cost of evaluating ML FFs lies in between empirical FFs and \textit{ab initio} reference calculations. In particular, the sGDML approach employed here is 5-10 orders of magnitude faster than \textit{ab initio} calculations and 2-3 orders of magnitude slower than classical FFs, this depends of course on the molecular system. For example, in the case of a CCSD(T)/cc-pVTZ calculation, the sGDML model can be up to 10$^7$ and 10$^9$ times faster for malondialdehyde and aspirin, respectively\cite{sGDMLsoftware2019}.

In the broadest sense, the challenge of accurately learning FFs is currently being addressed using two methods: Neural Networks (NN)~\cite{Behler2007,Behler2007b,Behler2011,Behler2011a,Behler2012,Behler2016,Gastegger2017,SchNetNIPS2017,SchNet2018} and kernel-based models~\cite{Bartok2013,Bartok2015_GAP,Ramprasad2015,Rupp2015,DeVita2015,eickenberg2018solid,Shapeev2017,Glielmo2017,noe2018,ResponseFieldAnatole2018,gdml,sgdml}. Both approaches can be constructed to employ energy and/or force information. Learning forces is advantageous for several reasons:
(i) FFs reconstruction in the force domain yield smoother PESs, eliminating artifacts due to somewhat conflicting requirement of simultaneously reproducing accurate energies and forces~\cite{SchNet2018,sGDMLsoftware2019}, the inherent uncertainty of the learning process, and using biased models that introduce unphysical approximations, e.g. atomic partitioning of the energy, (ii) obtaining energies from force models tends to diminish the noise in the prediction as a result of the integral operator, contrasting the behaviour of the forces generated by the gradient operator on energy models, and (iii) force models require smaller amounts of reference calculations to reach a desired accuracy~\cite{gdml}.
Such data efficiency arises not only due to the fact that each force data point carries 3$N$ components (where $N$ is the number of atoms) per reference calculation, but also because those components are orthogonal, thus providing complete information about the immediate local environment~\cite{LearnLinNIPS2002}.

Both NN and kernel-based methods can achieve formally any desired accuracy of predictions whenever a sufficiently large amount of training data is available. In contrast, when only 100s of data points are available, as in the case of high-level \textit{ab initio} data, the kernel methods usually offer a better reconstruction efficiency (with a unique and well-defined solution) as they make greater use of prior information.
Finally, it is important to emphasize the mandatory requirement of generating conservative ML-FFs, i.e. $\textbf{F}=-\nabla E$, to guarantee stable simulations.

The symmetric Gradient Domain Machine Learning (sGDML) FF~\cite{sgdml} retains all the advantages of kernel-based ML models which directly learn forces. 
In fact, training the sGDML model solely using forces, besides the availability of molecular energies, improves the learning process. Given that there is evidence that combining energies and forces in the loss function degrades the quality of the force prediction~\cite{sGDMLsoftware2019,SchNet2018}. The robustness of the method is explained by the fact that all atomic interactions are modelled globally, without resorting to an inherently non-unique partitioning into atom-wise, pairwise or many-body contributions. 
In the sGDML model: (i) Each FF model is explicitly constrained to be energy conserving, and
(ii) The model complexity is further reduced through the incorporation of molecular symmetries (i.e. rigid and fluxional) that are automatically extracted from the reference dataset. 
All these important properties contribute to the ability of sGDML to reconstruct complex PES for molecules of intermediate size from modest amounts of reference data, an unfeasible task for non-dedicated molecular FFs. In particular, the sGDML model enables the reconstruction of CCSD(T)-quality FFs from a limited amount ($\sim$ 100s) of reference molecular configurations~\cite{sgdml}.

In this article, we analyze some of the relevant quantum effects captured by the sGDML model while reconstructing the PES of small molecules. First, in section~\ref{sGDMLmodel}, we give a short introduction to the GDML framework and its symmetrized version, the sGDML model. In section~\ref{DiffML} a discussion regarding the advantages of gradient domain-based FFs is presented, as well as an analysis of dedicated vs. transferable FFs. Then, in section~\ref{MolPES}, we describe some of the quantum interactions described by the sGDML's reconstructed PES. In particular, we focus on three ubiquitous phenomena of general interest: section~\ref{MolPES}-A) lone-pairs and electrostatic interactions, section~\ref{MolPES}-B) intramolecular hydrogen bonds and proton transfer, and section~\ref{MolPES}-C) changes in atomic hybridization state and $n\to\pi^*$ interactions~\cite{n2pi_Aspirin2011,Blanco_n-pi2018}. Note that a qualitative description of these complex interactions by regular FF would require highly specialized models, while the sGDML model captures every interaction encoded in $-\textbf{F}_i=\langle \Psi^*|\partial \mathcal{H} /\partial \textbf{x}_i|\Psi \rangle$ with high accuracy. In the last section, we summarize our findings.

\begin{figure*}[htp]
\centering
\includegraphics[width=\textwidth]{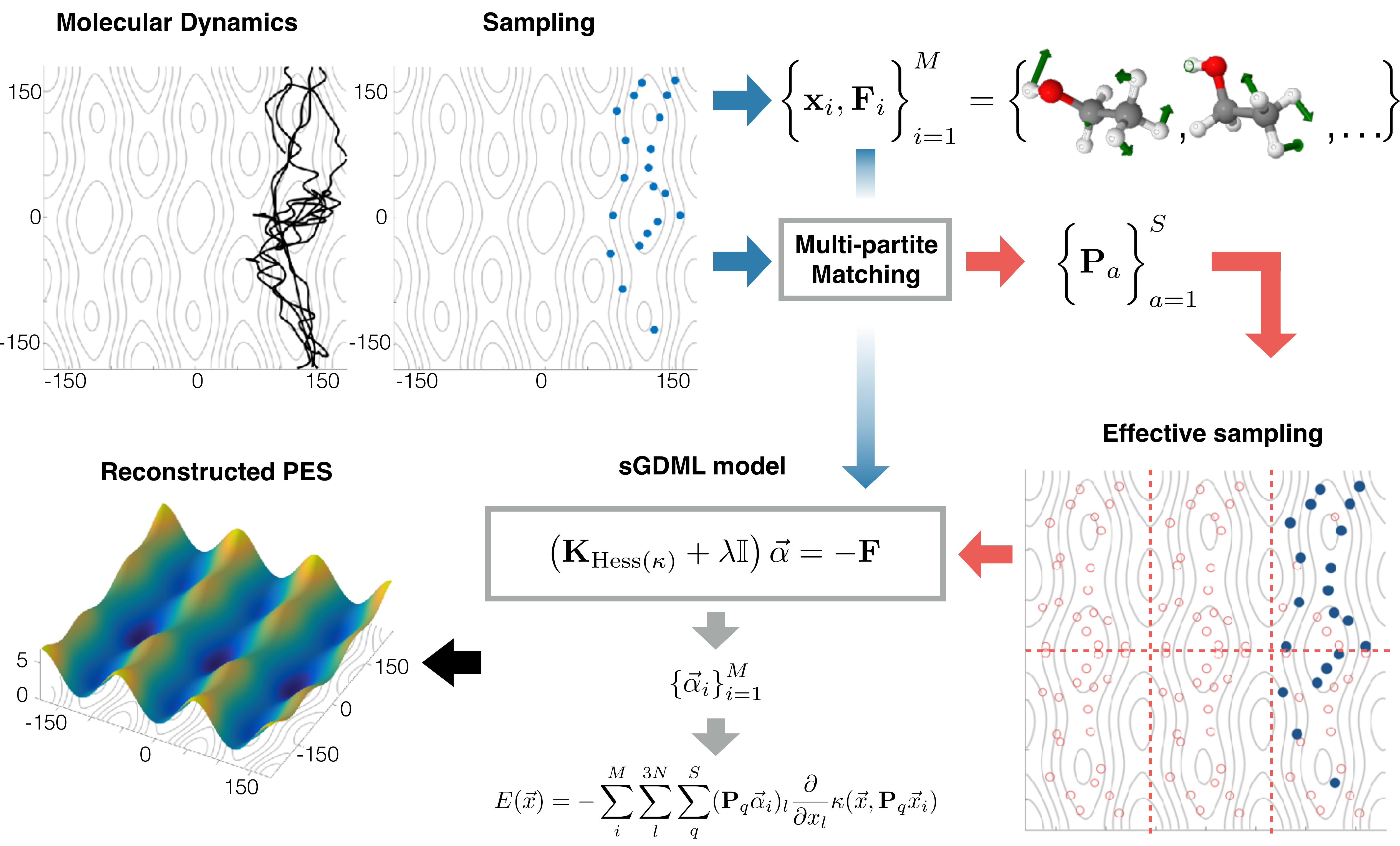}
\caption{Construction of the sGDML model. (1) The data used for training, \{\textbf{x}$_i$, \textbf{F}$_i$\}$_{i=1}^M$, is generated by random sampling of molecular dynamics trajectories (blue dots). The force on each atom is represented by a green arrow. (2) From the training set the permutational set of symmetries, \{\textbf{P}$_a$\}$_{a=1}^S$, are computed by the multi-partite matching approach. This effectively enhances the size of the training set by a factor $S$. (3) The force field is trained by solving the linear system for the parameters \{$\alpha_j$\}. The reconstructed potential-energy surface is obtained by analytically integrating the force model.}
\label{fig:FigModel}
\end{figure*}
%

\section{\texorpdfstring{\MakeLowercase{s}}~GDML model} \label{sGDMLmodel}

The sGDML model enables the direct efficient construction of dedicated FFs for flexible molecules from high-level \textit{ab initio} calculations. Unlike traditional FFs, it imposes no hypothesized analytical interaction models and thus in principle can model any physical interaction. Compared to other machine learning approaches, sGDML achieves high data efficiency through the incorporation of spatial and temporal physical symmetries. Global spatial symmetries include rotational and translational invariance of the energy, in addition to rigid and fluxional symmetries, which are recovered and enforced in an automatic data-driven manner. The homogeneity of time implies energy conservation, property that is enforced by learning in the gradient domain using as prior an analytically-integrable covariance function.

The latter is introduced as a linear operator constraint, by modeling the FF as the transformation of an underlying energy model. In particular, we train the gradient of a kernel ridge estimator on force labels $\mathbf{F}$, which -- by construction -- yields energy-conserving FFs that can be integrated to obtain the corresponding Born-Oppenheimer (BO) global potential-energy surface (PES) $V_{BO}$~\cite{gdml,sgdml}. Practically, this is achieved via the use of the Hessian matrix of a kernel $\kappa$ as the covariance structure to solve the normal equation of the ridge estimator in the gradient domain~\cite{gdml,sgdml}:
\begin{equation}
\left(\mathbf{K}_{\text{Hess}(\kappa)} + \lambda \mathbb{I}\right) \vec{\alpha}= \nabla V_{BO} = -\mathbf{F} \text{,}
\end{equation}
\noindent where $\mathbf{K}_{\text{Hess}(\kappa)}$ is the kernel matrix, $\lambda$ is the regularization parameter, $\mathbb{I}$ is the identity matrix, and $\vec{\alpha}$ are the parameter-vectors to train.

The sGDML model imposes additional permutational symmetry constraints on $\mathbf{K}_{\text{Hess}(\kappa)}$ to take advantage of PES redundancies due to rigid space group and fluxional symmetries~\cite{sgdml}. Typically, extracting those symmetries requires chemical and physical intuition about the system at hand, e.g. rotational barriers, which is impractical in a ML setting. Through a data-driven multi-partite matching approach (see Fig.~\ref{fig:FigModel}), we automate the discovery of permutation matrices $\mat{P}(\tau)$ corresponding to the index permutation $\tau$ from molecular dynamics simulations by realizing the assignment between adjacency matrices $(\mat{A})_{ij} = \|\vec{r}_i - \vec{r}_j\|$ of molecular graph pairs $G$ and $H$ in different energy states,
\begin{equation}
\operatorname*{arg\,min}_{\tau} \mathcal{L}(\tau) = \|\mat{P}(\tau)\mat{A}_G\mat{P}(\tau)\tran - \mat{A}_H\|^2 .
\label{eq:matching_objective}
\end{equation}

The resulting approximate pairwise matchings are subsequently synchronized using transitivity within the training set as the consistency criterion. A particular advantage of our solution is its ability to limit the symmetry recovery to energetically feasible permutational configurations, given that unfeasible permutation, e.g. the permutation of two random atoms, would not contribute any valuable information the symmetrized kernel and should not be considered. This severely reduces computational efforts in evaluating the model. Finally, the FF estimator trained on $M$ reference geometries, with $3N$ partial derivatives and $S$ symmetry transformations each, takes the form
\begin{equation}
\mat{\hat{f}_F}(\vec{x}) = \sum^M_{i} \sum^{3N}_{l} \sum^{S}_{q} (\mat{P}_q \vec{\alpha}_{i})_l  \frac{\partial}{\partial x_{l}} \nabla \kappa(\vec{x},\mat{P}_{q}\vec{x}_i)\,.
\label{eq:force_model}
\end{equation}

The corresponding energy predictor is obtained by simply integrating $\mat{\hat{f}_F}$ with respect to the Cartesian coordinates, 

\begin{equation}
-\hat{f}_E(\vec{x}) = \int \mat{\hat{f}_F} \cdot d\mat{x} = \sum^M_{i} \sum^{3N}_{l} \sum^{S}_{q} (\mat{P}_q \vec{\alpha}_{i})_l  \frac{\partial}{\partial x_{l}} \kappa(\vec{x},\mat{P}_{q}\vec{x}_i)\,.
\label{eq:force_model_e}
\end{equation}

Due to linearity of integration, the expression for the energy predictor is identical up to second derivative operator on the kernel function (see Fig.~\ref{fig:FigModel}). 
Figure~\ref{fig:FigModel} gives a general perspective of the sGDML model by summarizing the training process, from sampling the MD trajectory and extracting the permutational symmetries to solving the linear system and reconstructing the embedded PES in the data. 

The addition of spatial, temporal and permutational symmetry constraints leads to a gain in data efficiency of more than two orders of magnitude~\cite{sgdml}. 
Recently, we have systematically demonstrated that sGDML models trained on only few 100s of reference structures reconstruct molecular PESs with a mean average error of less that 0.06 kcal~$\text{mol}^{-1}$ for small molecules with up to 15 atoms and less than 0.16 kcal~$\text{mol}^{-1}$ for molecules as complex as aspirin, paracetamol, and azobenzene~\cite{sgdml} (See Table~\ref{tab:tab_dft2ccsd}, Tables S1 and S2). Hence, the explicit symmetrization incorporated in the GDML framework~\cite{gdml} results in robust learning models with the ability to preserve the complex subtleties encoded in the reference data.

The sGDML models for each molecule studied in this article were initially trained on DFT data at the generalized gradient approximation (GGA) level of theory with Perdew-Burke-Ernzerhof (PBE)~\cite{PBE1996} exchange-correlation functional and the Tkatchenko-Scheffler (TS) method~\cite{TS} to
account for van der Waals interactions. The training dataset was created by subsampling MD trajectories at constant temperature (500K) using the FHI-aims package~\cite{FHIaims2009}. In the case of keto-MDA, enol-MDA and ethanol we recomputed the training configurations using all-electron
CCSD(T), while in the case of Aspirin we used all-electron CCSD~\cite{psi42012,psi42017,psi42018}(see Supporting Information for further detains).

\section{Comparison of \texorpdfstring{\MakeLowercase{s}}~GDML to other ML-FF approaches} \label{DiffML}

\subsection{Force vs. energy model}

The unique approach used by the sGDML model contrasts other models that first develop an energy function and then get the forces by analytic differentiation~\cite{Alder1959,Rahman1964,Verlet1967,Rahman1971,EAM1984,Tersoff1988,PolarFF2006,TIP4P1983,TIP5P2000,AMBER1981,CHARMM1983,MMFF94,GROMOS2005,Behler2007,Behler2007b,Behler2011,Behler2011a,Behler2012,Behler2016,Gastegger2017,SchNetNIPS2017,SchNet2018,Bartok2013,Bartok2015_GAP,Ramprasad2015,Rupp2015,Glielmo2017,FCHL2018}. This is represented in the next diagram:

\begin{equation*}
  \begin{CD}
    &\text{Trained} & & \text{Derived} \\
    \text{sGDML}:&\mat{\hat{f}_F} @>>> \hat{f}_E = -\int \mat{\hat{f}_F} \cdot d\mat{x} \\
    \text{E-ML}: &\hat{f}_E  @>>> \mat{\hat{f}_F} = - \nabla \hat{f}_E
  \end{CD}
\end{equation*}

\noindent where the trained predictors and their post-training derived energy and forces are presented for the sGDML and an energy ML (E-ML) model, respectively. An interesting advantage of the sGDML over E-ML models is how the training error propagates to the derived quantities. Lets assume that the prediction errors associated with the models $\mat{\hat{f}_F}$ and $\hat{f}_E$ are $\gamma_F$ and $\gamma_E$, respectively. Then, from the discrete approximation of the integral and the derivative operator, we obtain that the error in the derived energy, $-\int \mat{\hat{f}_F} \cdot d\mat{x}$, is attenuated and given by $\sim\gamma_F\Delta x$ while the error in the derived forces, $-\nabla \hat{f}_E$, is amplified and given by $\sim\gamma_E / \Delta x$ (see Supporting Information for further detains). A direct implication of these results is that, as a whole, FFs based on E-ML are potentially less stable than gradient based FFs. Empirical evidence supporting these results as well as a proof from signal processing theory were published in the original GDML paper~\cite{gdml}.

Regarding the data efficiency of the sGDML, there is solid evidence from Gaussian processes (GPs) that learning linearizations of a function, e.g. gradients, is more informative than learning single points~\cite{LearnLinNIPS2002}. Such data efficiency has been systematically shown in the GDML framework~\cite{gdml,sgdml,sGDMLsoftware2019}. There is empirical evidence that more than $3N$ training data points would be needed in an E-ML model per each sample used in GDML to reach similar force accuracy~\cite{gdml}. Therefore, allowing to train molecular FFs using data from very accurate but computationally expensive reference methods, e.g. CCSD(T).

\subsection{Performance of using forces vs. forces+energies for training}

In the process of generating ML-FFs, the nature of the model, E-ML or gradient domain model, gives prior information regarding the problem to solve. This in the context of GPs would be equivalent to narrow the space of possible solutions. Then, a loss function is introduced in order to train the model by finding the best set of parameters that minimize such function (here presented without the regularization part):

\begin{equation*}
  \begin{CD}
    \text{Model}&\text{Loss function}\\
    \text{sGDML}:& loss_F = \sum_{i=1}^{M}{||\mat{\hat{f}_F}(\vec{x}_i)-\mat{F}_i||^2}\\
    \text{E-ML}:& loss_E = \sum_{i=1}^{M}{||\hat{f}_E(\vec{x}_i)-E_i||^2}
  \end{CD}
\end{equation*}

Using these loss functions would, in principle, give an optimal fitting respectively. There is the idea that such loss functions can be complemented by adding energy or force constraints, as they are often available in the reference data. In fact, several related works optimize a hybrid squared loss function of the form~\cite{SchNet2018}:

\begin{equation}
loss = \sum_{i=1}^{M}{\bigg \{||\mat{\hat{f}_F}(\vec{x}_i)-\mat{F}_i||^2+\eta ||\hat{f}_E(\vec{x}_i)-E_i||^2 \bigg \}},
\label{eq:fe_loss}
\end{equation}

\noindent where $\eta$ is a linear trade-off hyper-parameter which absorbs the differences in units and weights the force and energy contribution. By training a model using this loss function a somewhat conflicting optimization race between energies and forces is introduced.
Clear empirical evidence of this issue has been reported for NN-FFs~\cite{SchNet2018} and for the sGDML+E model~\cite{sGDMLsoftware2019} where in both cases the quality of the forces degrades by introducing energy constrains.

\subsection{Transferable and non-transferable models}

Recently, the idea of transferable or across chemical compound space ML-FFs has been under discussion but due to its complexity less progress has been achieved compared to dedicated ML-FFs. Transferable models can generate qualitatively good predictions simultaneously for different molecular systems~\cite{BoB2015,deltaML2015,Smith2017,amons2017}, but clearly cannot offer reliable results for PES reconstruction where energy prediction errors are often much larger than 1 kcal mol$^{-1}$~\cite{Smith2017}.
On the other hand, the accuracy achieved by state-of-the-art dedicated ML-FFs can even reach couple of \textit{calories} per mole in energy predictions using only a few hundreds reference calculations for training~\cite{sgdml}. In contrast, the above mentioned transferable model required $\sim$13 million data configurations for training. Furthermore, any gradient field generated from these models is, at the moment, not reliable because of such energy errors, which makes prohibitive to accurately capture physical interaction~\cite{Smith2017}. From this discussion it is apparent that transferable models cannot be used to study dynamical properties of molecules, a task easily  accomplished by dedicated FFs.

\begin{figure*}[htp]
\includegraphics[width=\textwidth]{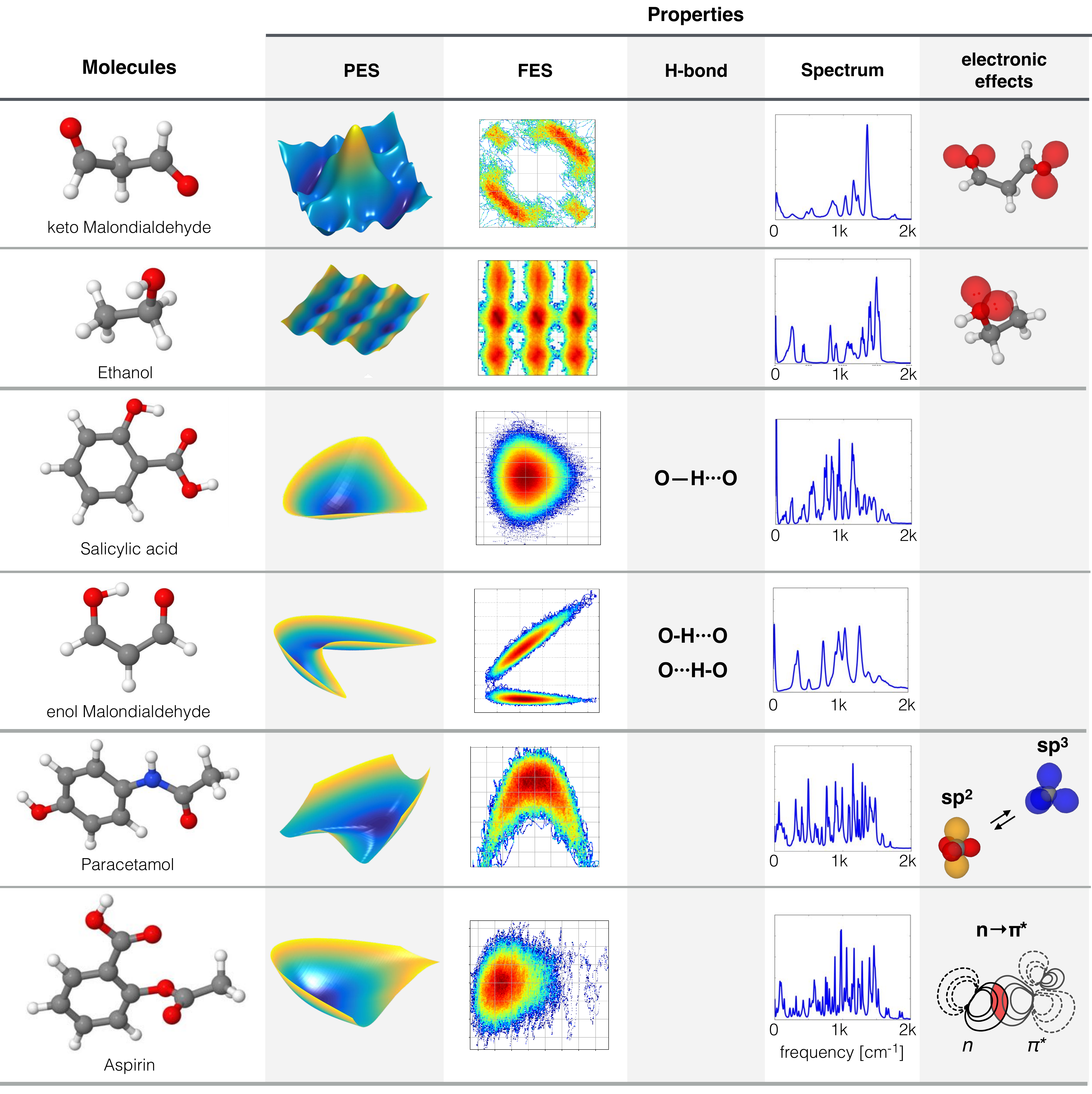}

\caption{Molecules under study and their properties. From left to right: List of molecules and their molecular structure, potential-energy surface along two relevant torsional degrees of freedom, free energy surface at 300 K, type of intramolecular hydrogen bonds (if applicable), vibrational spectrum at 300 K, and type of electronic effect to study (if applicable). The free energy scale from lowest (red) to higher (blue) is in $k_B$T. The last column shows the electron lone pairs in keto-MDA and ethanol, $sp^2 \to sp^3$ hybridization transition present in paracetamol and $n\to\pi^*$ interaction in aspirin\cite{n2pi_Aspirin2011}.}
\label{fig:FigContent}
\end{figure*}
\begin{figure*}[htp]
\includegraphics[width=\textwidth]{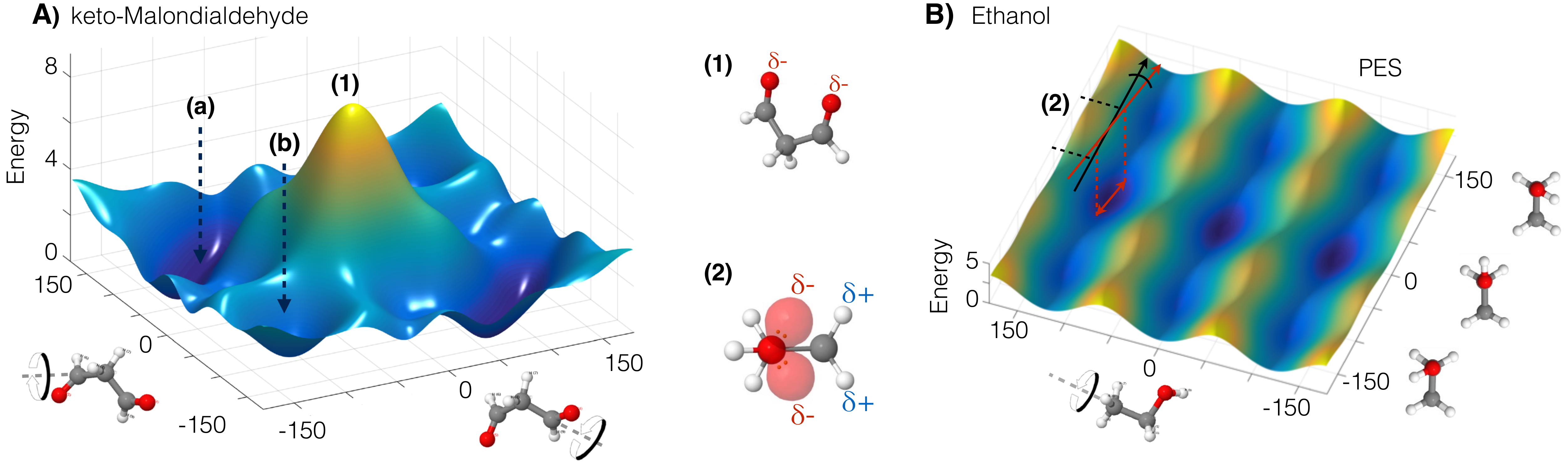}

\caption{Features of the PES. A) PES for keto--MDA. The structure (1) leads to a steep increase in energy due to the close distance between two negatively charged oxygen atoms. The regions (a) and (b) are the global and a local minima of keto--MDA. B) Ethanol's PES. Structure (2) shows the effect of oxygen's lone pair and the partial positive charges in methyl's hydrogen atoms and their coupling is represented by a red arrow in the PES of ethanol. Both PES were predicted using sGDML@CCSD(T) models~\cite{sgdml}.}
\label{fig:FigPES}
\end{figure*}
%

\section{Molecular potential-energy surfaces}  \label{MolPES}

In the framework of the BO approximation, $V_{BO}$ contains all the information necessary to describe the dynamics of a molecular system. All electronic quantum interactions are encoded in $V_{BO}$, but in practice it is not possible to expand $V_{BO}$ in different energetic contributions such as hydrogen bonding, electrostatics, dispersion interactions or other electronic effects.
Therefore, the intricate form of $V_{BO}$ resulting from an interplay between different quantum interactions, 
should be preserved in the reconstruction process. We will now demonstrate how sGDML is able to describe many complex features contained in the quantum-chemical conformational data.

%
\begin{table}[ht]
\caption{Accuracy of total energies for sGDML$@$DFT (using PBE+TS functional) and sGDML$@$CCSD(T) models on various molecular dynamics datasets. Energy errors are in kcal $\text{mol}^{-1}$. These results, with the exception of enol-MDA, were originally published in Ref.~\cite{sgdml}. All the models were trained using atomic forces for 1000 molecular conformations.}
\begin{threeparttable}
\label{tab:tab_dft2ccsd}
\setlength\extrarowheight{1pt}
\begin{ruledtabular}
\begin{tabular}{llrrr}
\multicolumn{1}{l}{Dataset} & \multicolumn{4}{l}{Energy Prediction}\\[2pt]
\cline{1-1} \cline{2-5}\\ [-2.3ex]
&\multicolumn{2}{l}{DFT}&\multicolumn{2}{l}{CCSD(T)}\\[2pt]
\cline{2-3} \cline{4-5}
 \\ [-2.3ex]
Molecule & MAE & RMSE & MAE & RMSE\\[1pt]
\hline \\ [-2.3ex]
keto--MDA & 
\ensuremath{0.10} &
\ensuremath{0.13} &
\ensuremath{0.06} &
\ensuremath{0.08}\\
Ethanol & 
\ensuremath{0.07} &
\ensuremath{0.09} &
\ensuremath{0.05} &
\ensuremath{0.07} \\
enol--MDA & 
\ensuremath{0.07} &
\ensuremath{0.09} &
\ensuremath{0.07} &
\ensuremath{0.07} \\
Aspirin & 
\ensuremath{0.19} &
\ensuremath{0.25} &
\ensuremath{0.16^{a}} &
\ensuremath{0.21^{a}}\\
\end{tabular}
\end{ruledtabular}
\begin{tablenotes}
      \item [a] CCSD
\end{tablenotes}
\end{threeparttable}
\end{table}
%

In practice, these important features or interactions (e.g. energy barriers or H-bond interactions) often come in the form of subtle variations in the $V_{BO}$ of less than 0.1 kcal~$\text{mol}^{-1}$, which is one order of magnitude lower than so-called chemical accuracy~\cite{sgdml}. For example, the relative stability of \textit{trans} and \textit{gauche} conformers of ethanol is within 0.1 kcal~$\text{mol}^{-1}$. Any model with an expected error above that threshold runs the risk of misrepresenting or even inverting this subtle energy difference, which will lead to incorrect occupation probabilities and hence qualitatively wrong dynamical properties.
The sGDML model has been shown to satisfy the stringent accuracy requirement of 0.1--0.2 kcal~$\text{mol}^{-1}$ for molecules with up to 15 atoms~\cite{sgdml}. Moreover, we found that using coupled-cluster reference data not only generates a more accurate description of the quantum system, but also improves the learning errors as shown in Table~\ref{tab:tab_dft2ccsd}. In this section, we exemplify the accuracy and insights obtained with sGDML with ubiquitous and challenging features of general interest in chemical physics: electron lone pairs, electrostatic interactions, intramolecular hydrogen bonds, proton tunneling effect, and other electronic effects (e.g. steric repulsion, change in the bond nature, and bonding--antibonding orbital interaction).
Figure~\ref{fig:FigContent} shows an overview of the different types of molecules and their reconstructed PES chosen to highlight the mentioned effects in this study.

\subsection{Electrostatic interactions and electron lone pairs} \label{QQint}

First, we focus our attention on electrostatic interactions, including atom--atom and lone-pair--atom interaction. 
Here, the concept of electron lone pairs play a central role; these are ubiquitous molecular features responsible for 
a wide variety of physical and chemical phenomena.
Lone pairs are valence electrons of an atom that are not shared with any other atom in a molecule.
Some examples of atoms in molecules that often present lone pairs are nitrogen and oxygen.
To illustrate the interactions induced by lone pairs, we will use the keto tautomer of malondialdehyde (keto--MDA) 
and ethanol molecule shown in the first two rows of Fig.~\ref{fig:FigContent}. 
These fluxional molecules have complex PES with a rich variety of physical phenomena (e.g. electrostatics and steric repulsion) for which the reconstruction process is not a trivial task (see Fig.~\ref{fig:FigPES}).

\subsubsection{Oxygen--oxygen atom repulsion in keto--MDA}

To illustrate the interatomic repulsion interaction we use the keto--MDA molecule, whose PES complexity is evident despite its small size (see Fig.~\ref{fig:FigPES}-A).
For example, the PES contains flat regions, which correspond to the global minima of the molecule (depicted in dark blue in Fig.~\ref{fig:FigPES}-A), but also display intricate pathways to move to local minima. Also, one can notice the sudden energy increase when the two oxygen atoms are in the closest configuration (structure (1) in Fig.~\ref{fig:FigPES}-A). 
 As mentioned before, the PES is the result of many complex interactions but certainly there are parts of the PES which 
 can be mainly ascribed to a particular phenomenon.
 This is the case for the yellow region in Fig.~\ref{fig:FigPES}-A, where the closeness between the two oxygen atoms suggests that the steep increase in the energy could be primarily attributed to the electrostatic repulsion between the lone pairs in each atom. 
 Additionally, we know that electron lone pair clouds have large spatial extent compared to shared electrons, therefore steric effects caused by electron cloud overlap could also be playing an important role in this region due to the close proximity between the two oxygen atoms ($r_{OO}\sim 2.6$ \AA).
From these two interactions, only the electrostatic contribution is roughly incorporated in regular FFs as constant point charges located on each atom. This greatly constrains their flexibility and reliability to describe complex interactions. 
Nonetheless, systematic studies of such interactions using the sGDML model could spawn new ideas regarding their integration into regular FFs and ultimately increase the predictive power of empirical FFs.

\begin{figure}[htp]
\centering
\includegraphics[width=1.0\columnwidth]{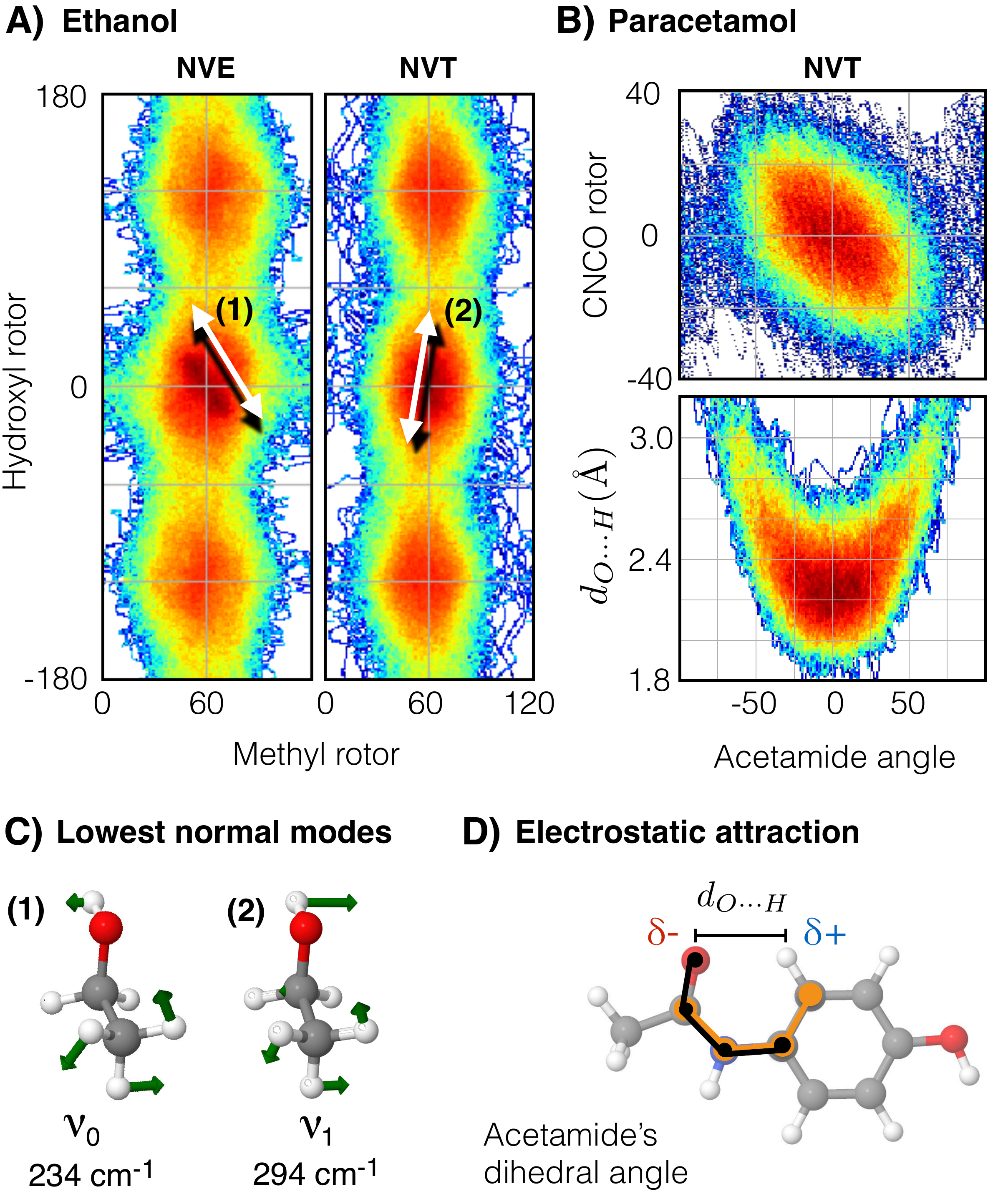}

\caption{Free energy for A) ethanol and B) paracetamol molecules at 300K using $F(T;x,y)=-k_B T \text{ln}[P(x,y)]$ where $P(x,y)$ is the sampling obtained from classical molecular dynamics. For ethanol two different ensembles are presented: NVE and NVT. C) Two lowest vibrational normal modes and their frequencies for \textit{trans} ethanol. The normal modes (1) and (2) are represented in the free energies A-NVE and A-NVT, respectively. D) Qualitative representation of the partial charges in paracetamol. The dihedral angle of the acetamide group respect to the benzene ring is represented in orange and the inner rotational degree of freedom CNCO is represented in black.}
\label{fig:FigSampling}
\end{figure}

 \subsubsection{Electron lone pairs in ethanol}

A particularly interesting case of strong effects of electron lone pairs is the ethanol molecule, where the lone pairs of the oxygen atom interact with the partially positive hydrogen atoms of the methyl group (structure (2) in Fig.~\ref{fig:FigPES}). This molecule has two rotors -- the hydroxyl and methyl groups -- as its main degrees of freedom. The PES for ethanol Fig.~\ref{fig:FigPES}-B exhibits a very subtle quasi-linear dependence between the dihedral angles of the methyl and hydroxyl functional groups in the \textit{trans} configuration (angle zero of the hydroxyl group in Fig.~\ref{fig:FigPES}-B) as shown by the red arrows in Fig.~\ref{fig:FigPES}-B. Such coupling is evident when analyzing the normal modes for this configuration, where the lowest two vibrations correspond to the aforementioned coupled motion (see also Fig.~\ref{fig:FigSampling}-C). 
The origin of this coupling can be understood by the electrostatic attraction between the lone pairs in the oxygen atom and the partially positively charged hydrogen atoms in the methyl rotor (structure (2) in Fig. \ref{fig:FigPES}). 
The correct description of this phenomenon within the reference data is crucial to obtain accurate physical properties of any molecular system with lone pairs~\cite{sgdml}. 
It is important to stress that an accurate and general description of lone pairs is a characteristic that goes beyond the capabilities of regular FFs~\cite{sgdml}. The handmade FFs that attempt to include lone pairs introduce explicit extra point charges~\cite{WaterReview2002}, which results in highly specialized models~\cite{AMBER-Hbond2017}.

\subsubsection{Dynamics of coupled rotors in ethanol}

The coupling between the hydroxyl and methyl rotors in ethanol manifests in the two lowest normal modes (1) and (2) shown in Fig.~\ref{fig:FigSampling}-C. The normal mode (1) corresponds to the direction indicated by the red arrow in the PES (Fig.~\ref{fig:FigPES}-B), while (2) moves perpendicularly. Here we analyze the dynamical implications of the coupling between lone pairs and the methyl rotor. 
By performing molecular dynamics simulations at 300K using the NVE and NVT ensembles, we have obtained different probability distributions and therefore different free energy surfaces (FES) as shown in Fig.~\ref{fig:FigSampling}-A. In both cases the FES considerably differs from its underlying PES in the \textit{trans} region. The FES$_{NVE}$ develops a deep minimum which traps the system into a state that boosts the occupation of the lowest vibrational mode, illustrated by (1) in Fig.~\ref{fig:FigSampling}-C. In contrast, FES$_{NVT}$ reverts the direction of the coupling as depicted in Fig.~\ref{fig:FigSampling}-A-(2), which in general promotes the occupation of both vibrational modes (1) and (2) in Fig.~\ref{fig:FigSampling}-C. In both cases, the FES shows an interesting and contrasting behavior that highlights the importance of the coupling between the two rotors in ethanol. A possible explanation for the difference between the two ensembles is that the NVT distributes the energy between all the molecular degrees of freedom in a more efficient way compared to NVE. The NVE ensemble relies only on the anharmonicities of the PES to redistribute the energy. Furthermore, the strong coupling between the two rotors suppresses the energy redistribution between vibrational normal modes in NVE.
It is clear that only FFs with an explicit description of lone pairs can show this coupling.

As shown here for ethanol and keto-MDA, the interactions involving electron lone pairs (e.g. electrostatic interactions, steric repulsion and $n\to\pi^*$ interactions) encoded in the PES play a fundamental role in the dynamics of the molecule and defining the free energy (Fig.~\ref{fig:FigContent}). This has direct implications on its thermodynamics and spectroscopic properties given that a different sampling of the PES re-weights and shifts the peaks in the vibrational spectrum. Therefore, since electron lone pairs participate in many other more complex interactions, their appropriate description is the first steps to generate reliable force fields.

\subsection{Intramolecular H--bond and proton transfer}
\begin{figure}[htp]
\centering
\includegraphics[width=1.0\columnwidth]{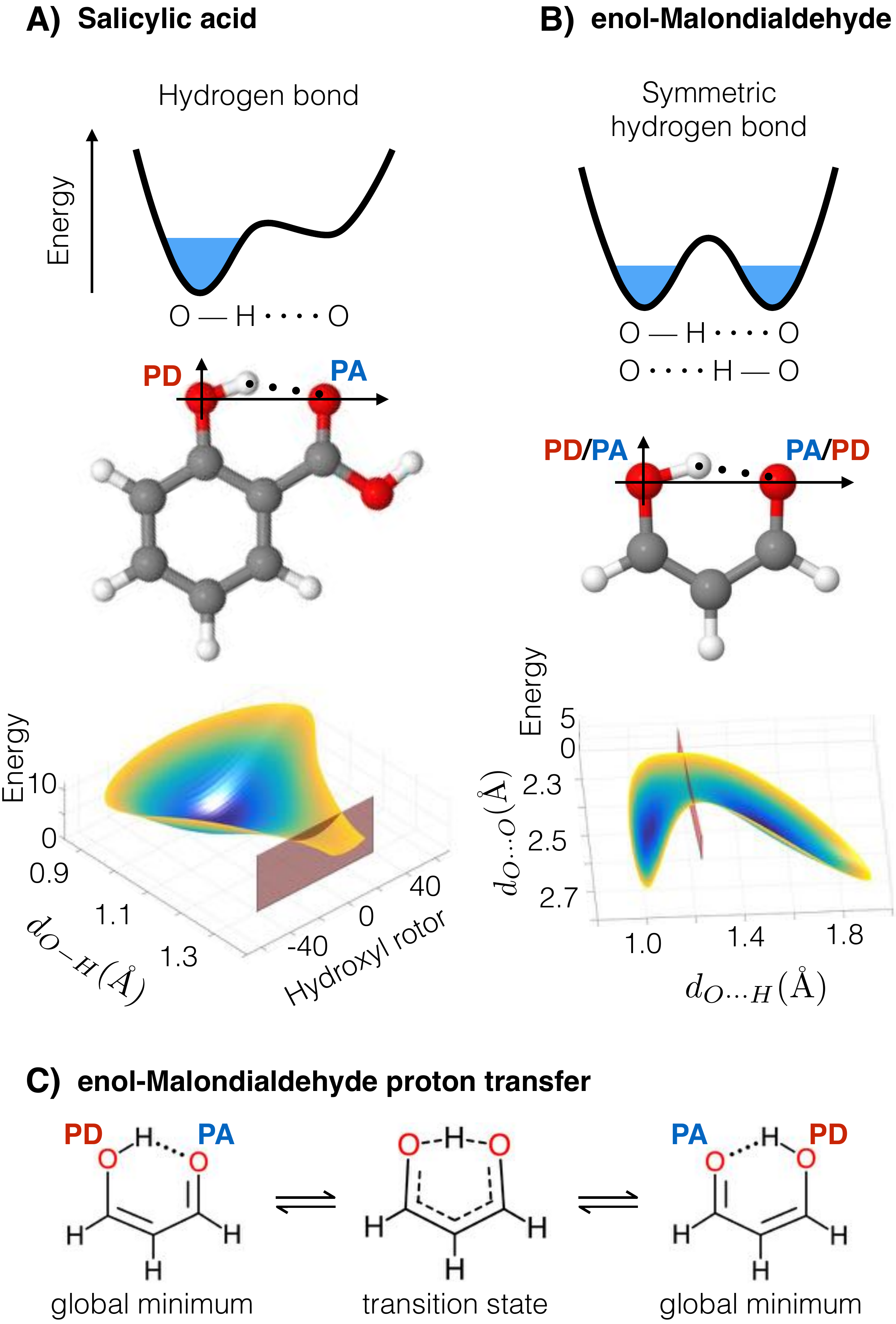}

\caption{Intramolecular hydrogen bond in A) salicylic acid and B) enol-Malondialdehyde. The top row shows schematically the type of H--bond in each case. In the middle row are the molecular structures and their respective proton reaction directions (from PD to PA). In the case of enol--MDA the PES is symmetric, then PD and PA are interchangeable. In the bottom row the PESs are shown where the red plane indicates the transition state of the proton. The energy at the transition states are $\sim$10 kcal~$\text{mol}^{-1}$ for salicylic acid and $\sim$4 kcal~$\text{mol}^{-1}$ for enol--MDA. In the transition state, the enol--MDA molecule has a $C_{2v}$ point symmetry, which the sGDML model exploits to increase the reconstruction accuracy.}
\label{fig:FigHB_PES}
\end{figure}

Another complex phenomenon accurately captured with the sGDML method is hydrogen bonding (H--bond). The \textit{intramolecular} H--bond is a subtle interaction, which dictates the dynamical behavior of many molecules~\cite{Scheiner2017}. It is responsible for very important molecular features such as the molecular structure and vibrational spectra, which result in macroscopic properties e.g. solubility and permeability~\cite{Kuhn2010}. Here, we will study two different types of H--bonds: standard donor--acceptor H--bond and the symmetric H--bond present in salicylic acid and the enol form of malondialdehyde (enol-MDA), respectively (Fig.~\ref{fig:FigHB_PES}). The symmetric H--bond allows proton tunneling to occur due to thermal fluctuations assisted by quantum nuclear effects to overcome the energetic barrier. In the case of a standard donor--acceptor H--bond, the proton is fixed to the proton donor (PD) while its dynamical behavior is strongly affected by the electron lone pair belonging to a neighboring atom (proton acceptor, PA). 
These two kinds of intramolecular H--bond appear very often in molecules and their presence can drastically change the physical properties of any molecule, as we show in this section for salicylic acid and enol-MDA molecules.

\subsubsection{Intramolecular H--bond}
\begin{figure}[htp]
\centering
\includegraphics[width=0.5\textwidth]{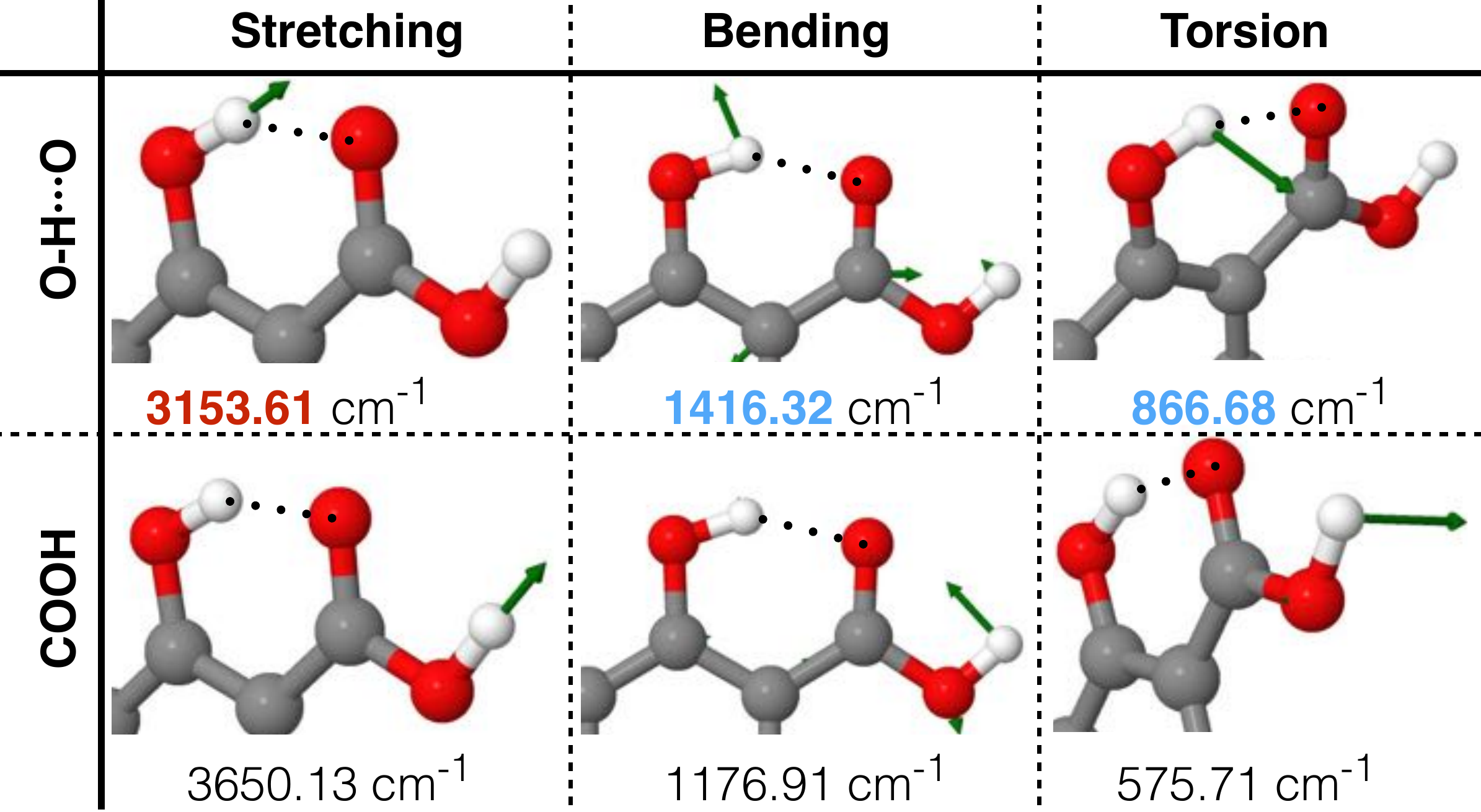}

\caption{Intramolecular H--bond in salicylic acid and its implications in the vibrational normal modes. The normal modes of the hydroxyl functional group involved in a H--bond (first row) and the reference ones in the carboxylic acid (COOH) functional group (second row).
}
\label{fig:FigSalicNM}
\end{figure}

We start by analyzing the salicylic acid molecule (Fig.~\ref{fig:FigHB_PES}-A). This molecule presents a standard donor--acceptor kind of H--bond between the hydroxyl and carboxylic acid groups. From the schematic representation in Fig.~\ref{fig:FigHB_PES}-A, we can see that the effect of the H--bond in this molecule consists in allowing the proton to stretch from the PD oxygen towards the PA oxygen. The middle point (transition state) between PD and PA is represented by a red plane in salicylic acid's PES in Fig.~\ref{fig:FigHB_PES}-A bottom. The energy necessary to reach this point is $\sim$10 kcal~$\text{mol}^{-1}$,  barely accessible at room temperature. We would also like to highlight the narrow structure of the reduced PES in the transition state (red plane in Fig.~\ref{fig:FigHB_PES}-A bottom), which gives us an idea of how directional the H--bond is. 
This directionality of the interaction changes the dynamics of the participating functional groups, which results in a characteristic red-shift in the stretching frequency of O--H induced by the H--bond~\cite{Hobza2002,Karpfen2009,Changwei2017}.
From a vibrational normal mode analysis on the sGDML reconstructed PES (see Fig.~\ref{fig:FigSalicNM}), we observe the red-shift of the O--H stretching frequency in the participating hydroxyl group. Furthermore, the H--bond also creates a blue shift in normal modes perpendicular to the H--bond (see Fig.~\ref{fig:FigSalicNM}), which is a direct evidence of a O--H$\cdots$O bond. The proper description of these molecular features will be directly displayed in spectroscopic properties such as IR and Raman spectrum which is often used to characterize molecular structures and their interactions.

\subsubsection{Proton transfer}

The second type of H--bond we analyze is the symmetric H--bond in enol--MDA, which exhibits a symmetric double-well reduced PES as schematically represented in Fig.~\ref{fig:FigHB_PES}-B. The energetic barrier separating the two minima is $\sim$4 kcal~$\text{mol}^{-1}$ which occurs when the interatomic distance between oxygen atoms is $d_{O\cdots O}=$2.38 \AA, this allows proton transfer between the two oxygen atoms even at room temperature. The transition state in the PES is shown by the red plane in Fig.~\ref{fig:FigHB_PES}-B bottom. 

The energetic barrier for the proton transfer has two possible contributions: Electron density rearrangement and quasi-aromaticity. 
From the schematic representation of the enol--MDA in Fig.~\ref{fig:FigHB_PES}-C, we see that going from the global minimum to the transition state entails a redistribution of the electron density. Therefore, the redistribution of $\pi$ electrons in the molecule induces an energetic penalty~\cite{ChangBook}, which contributes to the generation of a higher energy barrier. 
There is also evidence that molecules like enol--MDA behave as quasi-aromatic systems in the transition state~\cite{AromaOHN2017}. This phenomenon tends to stabilize the molecule in the transition state, which lowers the energetic barrier. 
Capturing these intricate and subtle quantum phenomena and their dynamics requires high-level quantum chemistry methods, with CCSD(T) being the only method that converges to the correct energetic barrier. We found that systematically increasing the amount of electron correlation energy in our calculations, the energy barrier decreases as $\sim$13 $\to$ $\sim$5 $\to$ $\sim$4 kcal~$\text{mol}^{-1}$ for HF $\to$ CCSD $\to$ CCSD(T), respectively. This result highlights the importance of the correlation energy in such complex phenomena as the H-bond interaction and proton transfer. 

The resulting free energy of the system obtained from MD simulations using sGDML@CCSD(T) at room temperature (see Fig.~\ref{fig:FigContent}) displays a very low proton transition rate between the two oxygen atoms but it is still accessible at room temperature.
This suggests that nuclear quantum effects would considerably increase the transition rate due to tunneling effects reshaping the FES and consequently its vibrational spectrum and thermodynamics.
In general, the local electron density delocalization induced by intramolecular H-bonds influence macroscopic properties such as solubility and permeability~\cite{Kuhn2010}, but a fundamentally different macroscopic implication of the symmetric H-bond is proton transport in extended systems like water. Therefore, the need of creating FFs capable of handling H-bonds in all of their flavors to accurately describe complex biological systems becomes obvious, and data-driven models enable a robust solution to this problem.

\subsection{Hybridization change $sp^2 \leftrightarrow sp^3$ and $n\to\pi^*$ interactions}

The two type of interactions mentioned in the previous subsections, electrostatic and H--bonds, are often approximately implemented in empirical FFs.
The advantages of flexible fully data-driven models are featured in describing any quantum interaction coming from $-\textbf{F}_i=\langle \Psi^*|\partial H /\partial \textbf{x}_i|\Psi \rangle$, without relying on prior knowledge of the phenomena or its connection to any classical electrodynamic or mechanical concepts.
To exemplify this, we consider the paracetamol and aspirin molecules. Their dynamics are strongly influenced by delocalized $\pi$ electrons and the $n\to\pi^*$ interaction~\cite{n-pi2017}, respectively. The quantitative description of these phenomena occurs naturally in our data-driven FF even when only a restricted amount of reference data is available. 
 
Hence, it is important to highlight that only now with the development of accurate ML-FFs trained on ab-initio data it is feasible to study in full extent the dynamical implication of such electronic effects at finite temperatures. 

\subsubsection{Hybridization state change in paracetamol}

Paracetamol is a molecule with a shallow global minimum consisting in a planar configuration (Fig.~\ref{fig:FigContent}) stabilized by being a conjugated system.
From Fig.~\ref{fig:FigContent} for paracetamol, a steep energy increase is evident as illustrated by yellow regions in the PES. This represents the breaking of the conjugated state, given that the nitrogen atom changes its hybridization state from $sp^2 \to sp^3$ producing an energetic penalty (see electronic effects column in Fig.~\ref{fig:FigContent}). 
Such electronic effects raise the energy, leading to an effectively inaccessible region in this direction of the PES. 
In fact, this region is hardly visited by MD simulations at 300 K, therefore it is not represented in the FES in Fig.~\ref{fig:FigSampling}-B.

Another important contribution to the planar structure of paracetamol is the electrostatic interaction between the lone pair on the carbonyl oxygen and the positively charged nearest hydrogen atom (see Fig.~\ref{fig:FigSampling}-D). We find a linear coupling between the acetamide main dihedral angle and the carbonyl dihedral angle, depicted in orange and black in Fig.~\ref{fig:FigSampling}-D, respectively.
The projected FES in these two variables shown in Fig.~\ref{fig:FigSampling}-B-top, reveals that near the global minimum
the system moves without altering the $d_{O\cdots H}$ distance since the internal dihedral angle CNCO flexes to follow the minimum free energy path. Certainly, paracetamol is a highly fluxional molecule containing four correlated rotors moving in a complex PES due to its electronic structure. 

\subsubsection{$n\to\pi^*$ interaction in Aspirin}

Another important electronic effect is the overlap between occupied (lone pair $n$) and antibonding ($\pi^*$) orbitals, this electronic effect is depicted in Fig.~\ref{fig:FigContent}. The aspirin molecule is a particularly interesting case in this regard given the dominant role of this interaction in its molecular behaviour. This crucial $n\to\pi^*$ attraction interaction is responsible for the binding between the ester and carbonyl groups, which dictates the structure of the global minimum. This effect is amplified at finite temperature given that thermal fluctuations enhance the overlap between the lone pair, $n$, in the carbonyl group and the antibonding orbital in the ester group, $\pi^*$~\cite{sgdml}. We have recently shown that the energy functional form of conventional FFs put into close contact four negatively charged oxygen atoms in aspirin; such strong charge repulsion leads to a misrepresentation of its PES~\cite{sgdml}. The sole incorporation of the missing lone pairs in all four closely interacting oxygen atoms and their directionality could greatly improve the results in regular FFs. 
In general, there are many other electronic effects (e.g. $n\to\sigma^*$ interactions~\cite{AceProNMe2016}, hyperconjugation, configuration dependent charge densities and Jahn--Teller effect~\cite{Exp03Spectro2017}) that are not explicitly incorporated in conventional FFs, nor captured by less robust ML-FF frameworks, which limits the reliability and predictive power of the dynamics. 
This rigorous requirement is justified by the increasing demand of computationally inexpensive and highly accurate PESs to interpret and obtain further insights into state-of-the-art spectroscopic experimental results~\cite{Fielicke2008,Fielicke2012,Exp01Spectro2010,Exp01Spectro2016,Exp01Spectro2017,Exp02Spectro2017,Exp03Spectro2017}.

In summary, we have analyzed a wide variety of energy landscapes reconstructed with high fidelity (see Fig.~\ref{fig:FigContent}). Being trained directly on molecular forces from \textit{ab initio} calculations, the generated PES encodes the broad range of fundamental interactions coming from the solution of the quantum-mechanical problem. This indicates that our model is a general ML approach capable of describing arbitrary interatomic interactions contained in the reference data.

\section{Conclusions}

We have presented the construction of molecular force fields using the symmetrized Gradient Domain Machine Learning model. This framework reconstructs high-dimensional manifold embedded in the training data from a few 100s of samples, allowing the use of highly-accurate \textit{ab initio} reference data such as the ``gold standard" CCSD(T) method. The flexibility of the sGDML model comes from its intrinsic nature of being a fully data-driven universal approximator, which grants the adaptability to describe any kind of quantum interaction.

This was demonstrated here by describing H-bonds, proton transfer, lone pairs, changes in hybridization states, steric repulsion, and $n\to\pi^*$ interactions and by obtaining insightful results from molecular dynamics simulations. 
From a careful analysis of the PES and MD simulations, we highlighted the importance of electron lone pairs in generating the strong coupling between the two rotors in ethanol and in the dynamics of keto-MDA.
On the other hand, the proper description of H-bonds revealed the proton dynamics in salicylic acid and enol-MDA molecules, yielding further understanding about its implications on the vibrational spectrum. Regarding electronic effects, the main contribution is that ML-FFs can be used as trustworthy tools able to describe non-trivial interactions. For example, from MD simulations we observed the $sp^2 \leftrightarrow sp^3$ hybridization change of the nitrogen atom in paracetamol which help us to understand better the strong consequences of breaking the conjugated molecular system, and that in aspirin the $n\to\pi^*$ interaction is enhanced at higher temperatures giving extra stability to the molecular global minima.

The main advantages of the sGDML model over other machine learning methods are copious: (i) it is highly data efficient, due to being trained in the gradient domain, (ii) it is robust due to modeling all atomic interactions globally, without any kind of inherent non-unique partitioning of the energy or force contributions, (iii) it uses energy conservation as a prior, therefore encoding this fundamental physical law in the core of every gradient-domain FF model, and (iv) it correctly represents spatial symmetries using explicit constraints that are automatically extracted from the data.

A number of challenges remain to be solved in order to extend the applicability of sGDML to larger systems. In spite of its many advantages, a global model imposes limits on the maximum molecule size, as well as the training set size. Overcoming this fundamental limitation without compromising its robustness calls for the introduction of a well-reasoned fragmentation scheme that divides the reconstruction problem into smaller independent subproblems without oversimplifying the nature of the interactions. A data-driven approach could achieve this task in a way that is tailored to preserving the intricate phenomena studied in this article, as opposed to applying general coarse-graining techniques. Such an approach will benefit from the explicit knowledge about fluxional symmetries within the system, which our algorithm is already able to extract.
In its current formulation, the sGDML model captures different interaction scales, with no need to separating them. Nevertheless, an explicit decoupling of long-range interactions could be a new avenue to further increase data efficiency on the way to increasingly larger and complex molecules. 

\section{Supplementary Material}
See supplementary material for additional tables with prediction accuracy for total energies and forces from DFT and CCSD(T) data. Also, some supplementary notes regarding reference data generation, molecular dynamics details, and error propagation. 

The sGDML code and documentation is available at
http://quantum-machine.org/gdml/.

\section{ACKNOWLEDGEMENTS}

We thank Dr. M. Gastegger for helpful discussions. S.C., A.T., and K.-R.M. thank the Deutsche Forschungsgemeinschaft (project MU 987/20-1) for funding this work. A.T. is funded by the European Research Council with ERC-CoG grant BeStMo. KRM acknowledges partial support by BMBF (BZML and BBDC) as well as by the Institute of Information \& Communications Technology Planning \& Evaluation (IITP) grant funded by the Korea government (No. 2017-0-00451). Part of this research was performed while the authors were visiting the Institute for Pure and Applied Mathematics, which is supported by the NSF.


\bibliography{main}

\begin{thebibliography}{100}%
\makeatletter
\providecommand \@ifxundefined [1]{%
 \@ifx{#1\undefined}
}%
\providecommand \@ifnum [1]{%
 \ifnum #1\expandafter \@firstoftwo
 \else \expandafter \@secondoftwo
 \fi
}%
\providecommand \@ifx [1]{%
 \ifx #1\expandafter \@firstoftwo
 \else \expandafter \@secondoftwo
 \fi
}%
\providecommand \natexlab [1]{#1}%
\providecommand \enquote  [1]{``#1''}%
\providecommand \bibnamefont  [1]{#1}%
\providecommand \bibfnamefont [1]{#1}%
\providecommand \citenamefont [1]{#1}%
\providecommand \href@noop [0]{\@secondoftwo}%
\providecommand \href [0]{\begingroup \@sanitize@url \@href}%
\providecommand \@href[1]{\@@startlink{#1}\@@href}%
\providecommand \@@href[1]{\endgroup#1\@@endlink}%
\providecommand \@sanitize@url [0]{\catcode `\\12\catcode `\$12\catcode
  `\&12\catcode `\#12\catcode `\^12\catcode `\_12\catcode `\%12\relax}%
\providecommand \@@startlink[1]{}%
\providecommand \@@endlink[0]{}%
\providecommand \url  [0]{\begingroup\@sanitize@url \@url }%
\providecommand \@url [1]{\endgroup\@href {#1}{\urlprefix }}%
\providecommand \urlprefix  [0]{URL }%
\providecommand \Eprint [0]{\href }%
\providecommand \doibase [0]{http://dx.doi.org/}%
\providecommand \selectlanguage [0]{\@gobble}%
\providecommand \bibinfo  [0]{\@secondoftwo}%
\providecommand \bibfield  [0]{\@secondoftwo}%
\providecommand \translation [1]{[#1]}%
\providecommand \BibitemOpen [0]{}%
\providecommand \bibitemStop [0]{}%
\providecommand \bibitemNoStop [0]{.\EOS\space}%
\providecommand \EOS [0]{\spacefactor3000\relax}%
\providecommand \BibitemShut  [1]{\csname bibitem#1\endcsname}%
\let\auto@bib@innerbib\@empty
\bibitem [{\citenamefont {Alder}\ and\ \citenamefont
  {Wainwright}(1959)}]{Alder1959}%
  \BibitemOpen
  \bibfield  {author} {\bibinfo {author} {\bibfnamefont {B.~J.}\ \bibnamefont
  {Alder}}\ and\ \bibinfo {author} {\bibfnamefont {T.~E.}\ \bibnamefont
  {Wainwright}},\ }\href {\doibase 10.1063/1.1730376} {\bibfield  {journal}
  {\bibinfo  {journal} {J. Chem. Phys.}\ }\textbf {\bibinfo {volume} {31}},\
  \bibinfo {pages} {459} (\bibinfo {year} {1959})}\BibitemShut {NoStop}%
\bibitem [{\citenamefont {Rahman}(1964)}]{Rahman1964}%
  \BibitemOpen
  \bibfield  {author} {\bibinfo {author} {\bibfnamefont {A.}~\bibnamefont
  {Rahman}},\ }\href {\doibase 10.1103/PhysRev.136.A405} {\bibfield  {journal}
  {\bibinfo  {journal} {Phys. Rev.}\ }\textbf {\bibinfo {volume} {136}},\
  \bibinfo {pages} {A405} (\bibinfo {year} {1964})}\BibitemShut {NoStop}%
\bibitem [{\citenamefont {Verlet}(1967)}]{Verlet1967}%
  \BibitemOpen
  \bibfield  {author} {\bibinfo {author} {\bibfnamefont {L.}~\bibnamefont
  {Verlet}},\ }\href {\doibase 10.1103/PhysRev.159.98} {\bibfield  {journal}
  {\bibinfo  {journal} {Phys. Rev.}\ }\textbf {\bibinfo {volume} {159}},\
  \bibinfo {pages} {98} (\bibinfo {year} {1967})}\BibitemShut {NoStop}%
\bibitem [{\citenamefont {Rahman}\ and\ \citenamefont
  {Stillinger}(1971)}]{Rahman1971}%
  \BibitemOpen
  \bibfield  {author} {\bibinfo {author} {\bibfnamefont {A.}~\bibnamefont
  {Rahman}}\ and\ \bibinfo {author} {\bibfnamefont {F.~H.}\ \bibnamefont
  {Stillinger}},\ }\href {\doibase 10.1063/1.1676585} {\bibfield  {journal}
  {\bibinfo  {journal} {J. Chem. Phys.}\ }\textbf {\bibinfo {volume} {55}},\
  \bibinfo {pages} {3336} (\bibinfo {year} {1971})}\BibitemShut {NoStop}%
\bibitem [{\citenamefont {Daw}\ and\ \citenamefont {Baskes}(1984)}]{EAM1984}%
  \BibitemOpen
  \bibfield  {author} {\bibinfo {author} {\bibfnamefont {M.~S.}\ \bibnamefont
  {Daw}}\ and\ \bibinfo {author} {\bibfnamefont {M.~I.}\ \bibnamefont
  {Baskes}},\ }\href {\doibase 10.1103/PhysRevB.29.6443} {\bibfield  {journal}
  {\bibinfo  {journal} {Phys. Rev. B}\ }\textbf {\bibinfo {volume} {29}},\
  \bibinfo {pages} {6443} (\bibinfo {year} {1984})}\BibitemShut {NoStop}%
\bibitem [{\citenamefont {Tersoff}(1988)}]{Tersoff1988}%
  \BibitemOpen
  \bibfield  {author} {\bibinfo {author} {\bibfnamefont {J.}~\bibnamefont
  {Tersoff}},\ }\href {\doibase 10.1103/PhysRevB.37.6991} {\bibfield  {journal}
  {\bibinfo  {journal} {Phys. Rev. B}\ }\textbf {\bibinfo {volume} {37}},\
  \bibinfo {pages} {6991} (\bibinfo {year} {1988})}\BibitemShut {NoStop}%
\bibitem [{\citenamefont {Warshel}\ \emph {et~al.}(2006)\citenamefont
  {Warshel}, \citenamefont {Sharma}, \citenamefont {Kato},\ and\ \citenamefont
  {Parson}}]{PolarFF2006}%
  \BibitemOpen
  \bibfield  {author} {\bibinfo {author} {\bibfnamefont {A.}~\bibnamefont
  {Warshel}}, \bibinfo {author} {\bibfnamefont {P.~K.}\ \bibnamefont {Sharma}},
  \bibinfo {author} {\bibfnamefont {M.}~\bibnamefont {Kato}}, \ and\ \bibinfo
  {author} {\bibfnamefont {W.~W.}\ \bibnamefont {Parson}},\ }\href {\doibase
  https://doi.org/10.1016/j.bbapap.2006.08.007} {\bibfield  {journal} {\bibinfo
   {journal} {Biochim. Biophys. Acta, Proteins Proteomics}\ }\textbf {\bibinfo
  {volume} {1764}},\ \bibinfo {pages} {1647 } (\bibinfo {year}
  {2006})}\BibitemShut {NoStop}%
\bibitem [{\citenamefont {Jorgensen}\ \emph {et~al.}(1983)\citenamefont
  {Jorgensen}, \citenamefont {Chandrasekhar}, \citenamefont {Madura},
  \citenamefont {Impey},\ and\ \citenamefont {Klein}}]{TIP4P1983}%
  \BibitemOpen
  \bibfield  {author} {\bibinfo {author} {\bibfnamefont {W.~L.}\ \bibnamefont
  {Jorgensen}}, \bibinfo {author} {\bibfnamefont {J.}~\bibnamefont
  {Chandrasekhar}}, \bibinfo {author} {\bibfnamefont {J.~D.}\ \bibnamefont
  {Madura}}, \bibinfo {author} {\bibfnamefont {R.~W.}\ \bibnamefont {Impey}}, \
  and\ \bibinfo {author} {\bibfnamefont {M.~L.}\ \bibnamefont {Klein}},\ }\href
  {\doibase 10.1063/1.445869} {\bibfield  {journal} {\bibinfo  {journal} {J.
  Chem. Phys.}\ }\textbf {\bibinfo {volume} {79}},\ \bibinfo {pages} {926}
  (\bibinfo {year} {1983})}\BibitemShut {NoStop}%
\bibitem [{\citenamefont {Mahoney}\ and\ \citenamefont
  {Jorgensen}(2000)}]{TIP5P2000}%
  \BibitemOpen
  \bibfield  {author} {\bibinfo {author} {\bibfnamefont {M.~W.}\ \bibnamefont
  {Mahoney}}\ and\ \bibinfo {author} {\bibfnamefont {W.~L.}\ \bibnamefont
  {Jorgensen}},\ }\href {\doibase 10.1063/1.481505} {\bibfield  {journal}
  {\bibinfo  {journal} {J. Chem. Phys.}\ }\textbf {\bibinfo {volume} {112}},\
  \bibinfo {pages} {8910} (\bibinfo {year} {2000})}\BibitemShut {NoStop}%
\bibitem [{\citenamefont {Weiner}\ and\ \citenamefont
  {Kollman}(1981)}]{AMBER1981}%
  \BibitemOpen
  \bibfield  {author} {\bibinfo {author} {\bibfnamefont {P.~K.}\ \bibnamefont
  {Weiner}}\ and\ \bibinfo {author} {\bibfnamefont {P.~A.}\ \bibnamefont
  {Kollman}},\ }\href {\doibase 10.1002/jcc.540020311} {\bibfield  {journal}
  {\bibinfo  {journal} {J. Comput. Chem.}\ }\textbf {\bibinfo {volume} {2}},\
  \bibinfo {pages} {287} (\bibinfo {year} {1981})}\BibitemShut {NoStop}%
\bibitem [{\citenamefont {Brooks}\ \emph {et~al.}(1983)\citenamefont {Brooks},
  \citenamefont {Bruccoleri}, \citenamefont {Olafson}, \citenamefont {States},
  \citenamefont {Swaminathan},\ and\ \citenamefont {Karplus}}]{CHARMM1983}%
  \BibitemOpen
  \bibfield  {author} {\bibinfo {author} {\bibfnamefont {B.~R.}\ \bibnamefont
  {Brooks}}, \bibinfo {author} {\bibfnamefont {R.~E.}\ \bibnamefont
  {Bruccoleri}}, \bibinfo {author} {\bibfnamefont {B.~D.}\ \bibnamefont
  {Olafson}}, \bibinfo {author} {\bibfnamefont {D.~J.}\ \bibnamefont {States}},
  \bibinfo {author} {\bibfnamefont {S.}~\bibnamefont {Swaminathan}}, \ and\
  \bibinfo {author} {\bibfnamefont {M.}~\bibnamefont {Karplus}},\ }\href
  {\doibase 10.1002/jcc.540040211} {\bibfield  {journal} {\bibinfo  {journal}
  {J. Comput. Chem.}\ }\textbf {\bibinfo {volume} {4}},\ \bibinfo {pages} {187}
  (\bibinfo {year} {1983})}\BibitemShut {NoStop}%
\bibitem [{\citenamefont {Halgren}(1996)}]{MMFF94}%
  \BibitemOpen
  \bibfield  {author} {\bibinfo {author} {\bibfnamefont {T.~A.}\ \bibnamefont
  {Halgren}},\ }\href {\doibase
  10.1002/(SICI)1096-987X(199604)17:5/6<490::AID-JCC1>3.0.CO;2-P} {\bibfield
  {journal} {\bibinfo  {journal} {J. Comput. Chem.}\ }\textbf {\bibinfo
  {volume} {17}},\ \bibinfo {pages} {490} (\bibinfo {year} {1996})}\BibitemShut
  {NoStop}%
\bibitem [{\citenamefont {Soares}\ \emph {et~al.}(2005)\citenamefont {Soares},
  \citenamefont {H\"unenberger}, \citenamefont {Kastenholz}, \citenamefont
  {Kr\"autler}, \citenamefont {Lenz}, \citenamefont {Lins}, \citenamefont
  {Oostenbrink},\ and\ \citenamefont {van Gunsteren}}]{GROMOS2005}%
  \BibitemOpen
  \bibfield  {author} {\bibinfo {author} {\bibfnamefont {T.~A.}\ \bibnamefont
  {Soares}}, \bibinfo {author} {\bibfnamefont {P.~H.}\ \bibnamefont
  {H\"unenberger}}, \bibinfo {author} {\bibfnamefont {M.~A.}\ \bibnamefont
  {Kastenholz}}, \bibinfo {author} {\bibfnamefont {V.}~\bibnamefont
  {Kr\"autler}}, \bibinfo {author} {\bibfnamefont {T.}~\bibnamefont {Lenz}},
  \bibinfo {author} {\bibfnamefont {R.~D.}\ \bibnamefont {Lins}}, \bibinfo
  {author} {\bibfnamefont {C.}~\bibnamefont {Oostenbrink}}, \ and\ \bibinfo
  {author} {\bibfnamefont {W.~F.}\ \bibnamefont {van Gunsteren}},\ }\href
  {\doibase 10.1002/jcc.20193} {\bibfield  {journal} {\bibinfo  {journal} {J.
  Comput. Chem.}\ }\textbf {\bibinfo {volume} {26}},\ \bibinfo {pages} {725}
  (\bibinfo {year} {2005})}\BibitemShut {NoStop}%
\bibitem [{\citenamefont {Rupp}\ \emph {et~al.}(2012)\citenamefont {Rupp},
  \citenamefont {Tkatchenko}, \citenamefont {M{\"{u}}ller},\ and\ \citenamefont
  {von Lilienfeld}}]{Rupp2012}%
  \BibitemOpen
  \bibfield  {author} {\bibinfo {author} {\bibfnamefont {M.}~\bibnamefont
  {Rupp}}, \bibinfo {author} {\bibfnamefont {A.}~\bibnamefont {Tkatchenko}},
  \bibinfo {author} {\bibfnamefont {K.-R.}\ \bibnamefont {M{\"{u}}ller}}, \
  and\ \bibinfo {author} {\bibfnamefont {O.~A.}\ \bibnamefont {von
  Lilienfeld}},\ }\href {\doibase 10.1103/PhysRevLett.108.058301} {\bibfield
  {journal} {\bibinfo  {journal} {Phys. Rev. Lett.}\ }\textbf {\bibinfo
  {volume} {108}},\ \bibinfo {pages} {58301} (\bibinfo {year}
  {2012})}\BibitemShut {NoStop}%
\bibitem [{\citenamefont {Bart{\'{o}}k}\ \emph {et~al.}(2010)\citenamefont
  {Bart{\'{o}}k}, \citenamefont {Payne}, \citenamefont {Kondor},\ and\
  \citenamefont {Cs{\'{a}}nyi}}]{Bartok2010}%
  \BibitemOpen
  \bibfield  {author} {\bibinfo {author} {\bibfnamefont {A.~P.}\ \bibnamefont
  {Bart{\'{o}}k}}, \bibinfo {author} {\bibfnamefont {M.~C.}\ \bibnamefont
  {Payne}}, \bibinfo {author} {\bibfnamefont {R.}~\bibnamefont {Kondor}}, \
  and\ \bibinfo {author} {\bibfnamefont {G.}~\bibnamefont {Cs{\'{a}}nyi}},\
  }\href {\doibase 10.1103/PhysRevLett.104.136403} {\bibfield  {journal}
  {\bibinfo  {journal} {Phys. Rev. Lett.}\ }\textbf {\bibinfo {volume} {104}},\
  \bibinfo {pages} {136403} (\bibinfo {year} {2010})}\BibitemShut {NoStop}%
\bibitem [{\citenamefont {Hansen}\ \emph {et~al.}(2013)\citenamefont {Hansen},
  \citenamefont {Montavon}, \citenamefont {Biegler}, \citenamefont {Fazli},
  \citenamefont {Rupp}, \citenamefont {Scheffler}, \citenamefont {von
  Lilienfeld}, \citenamefont {Tkatchenko},\ and\ \citenamefont
  {M{\"{u}}ller}}]{Hansen2013}%
  \BibitemOpen
  \bibfield  {author} {\bibinfo {author} {\bibfnamefont {K.}~\bibnamefont
  {Hansen}}, \bibinfo {author} {\bibfnamefont {G.}~\bibnamefont {Montavon}},
  \bibinfo {author} {\bibfnamefont {F.}~\bibnamefont {Biegler}}, \bibinfo
  {author} {\bibfnamefont {S.}~\bibnamefont {Fazli}}, \bibinfo {author}
  {\bibfnamefont {M.}~\bibnamefont {Rupp}}, \bibinfo {author} {\bibfnamefont
  {M.}~\bibnamefont {Scheffler}}, \bibinfo {author} {\bibfnamefont {O.~A.}\
  \bibnamefont {von Lilienfeld}}, \bibinfo {author} {\bibfnamefont
  {A.}~\bibnamefont {Tkatchenko}}, \ and\ \bibinfo {author} {\bibfnamefont
  {K.-R.}\ \bibnamefont {M{\"{u}}ller}},\ }\href {\doibase 10.1021/ct400195d}
  {\bibfield  {journal} {\bibinfo  {journal} {J. Chem. Theory Comput.}\
  }\textbf {\bibinfo {volume} {9}},\ \bibinfo {pages} {3404} (\bibinfo {year}
  {2013})}\BibitemShut {NoStop}%
\bibitem [{\citenamefont {Bart{\'{o}}k}\ and\ \citenamefont
  {Cs{\'{a}}nyi}(2015)}]{Bartok2015_GAP}%
  \BibitemOpen
  \bibfield  {author} {\bibinfo {author} {\bibfnamefont {A.~P.}\ \bibnamefont
  {Bart{\'{o}}k}}\ and\ \bibinfo {author} {\bibfnamefont {G.}~\bibnamefont
  {Cs{\'{a}}nyi}},\ }\href {\doibase 10.1002/qua.24927} {\bibfield  {journal}
  {\bibinfo  {journal} {Int. J. Quantum Chem.}\ }\textbf {\bibinfo {volume}
  {115}},\ \bibinfo {pages} {1051} (\bibinfo {year} {2015})}\BibitemShut
  {NoStop}%
\bibitem [{\citenamefont {Hansen}\ \emph
  {et~al.}(2015{\natexlab{a}})\citenamefont {Hansen}, \citenamefont {Biegler},
  \citenamefont {Ramakrishnan}, \citenamefont {Pronobis}, \citenamefont {von
  Lilienfeld}, \citenamefont {M{\"{u}}ller},\ and\ \citenamefont
  {Tkatchenko}}]{Hansen2015}%
  \BibitemOpen
  \bibfield  {author} {\bibinfo {author} {\bibfnamefont {K.}~\bibnamefont
  {Hansen}}, \bibinfo {author} {\bibfnamefont {F.}~\bibnamefont {Biegler}},
  \bibinfo {author} {\bibfnamefont {R.}~\bibnamefont {Ramakrishnan}}, \bibinfo
  {author} {\bibfnamefont {W.}~\bibnamefont {Pronobis}}, \bibinfo {author}
  {\bibfnamefont {O.~A.}\ \bibnamefont {von Lilienfeld}}, \bibinfo {author}
  {\bibfnamefont {K.-R.}\ \bibnamefont {M{\"{u}}ller}}, \ and\ \bibinfo
  {author} {\bibfnamefont {A.}~\bibnamefont {Tkatchenko}},\ }\href {\doibase
  10.1021/acs.jpclett.5b00831} {\bibfield  {journal} {\bibinfo  {journal} {J.
  Phys. Chem. Lett.}\ }\textbf {\bibinfo {volume} {6}},\ \bibinfo {pages}
  {2326} (\bibinfo {year} {2015}{\natexlab{a}})}\BibitemShut {NoStop}%
\bibitem [{\citenamefont {Rupp}\ \emph {et~al.}(2015)\citenamefont {Rupp},
  \citenamefont {Ramakrishnan},\ and\ \citenamefont {von
  Lilienfeld}}]{Rupp2015}%
  \BibitemOpen
  \bibfield  {author} {\bibinfo {author} {\bibfnamefont {M.}~\bibnamefont
  {Rupp}}, \bibinfo {author} {\bibfnamefont {R.}~\bibnamefont {Ramakrishnan}},
  \ and\ \bibinfo {author} {\bibfnamefont {O.~A.}\ \bibnamefont {von
  Lilienfeld}},\ }\href {\doibase 10.1021/acs.jpclett.5b01456} {\bibfield
  {journal} {\bibinfo  {journal} {J. Phys. Chem. Lett.}\ }\textbf {\bibinfo
  {volume} {6}},\ \bibinfo {pages} {3309} (\bibinfo {year} {2015})}\BibitemShut
  {NoStop}%
\bibitem [{\citenamefont {De}\ \emph {et~al.}(2016)\citenamefont {De},
  \citenamefont {Bartok}, \citenamefont {Cs{\'{a}}nyi},\ and\ \citenamefont
  {Ceriotti}}]{Ceriotti2016}%
  \BibitemOpen
  \bibfield  {author} {\bibinfo {author} {\bibfnamefont {S.}~\bibnamefont
  {De}}, \bibinfo {author} {\bibfnamefont {A.~P.}\ \bibnamefont {Bartok}},
  \bibinfo {author} {\bibfnamefont {G.}~\bibnamefont {Cs{\'{a}}nyi}}, \ and\
  \bibinfo {author} {\bibfnamefont {M.}~\bibnamefont {Ceriotti}},\ }\href
  {\doibase 10.1039/C6CP00415F} {\bibfield  {journal} {\bibinfo  {journal}
  {Phys. Chem. Chem. Phys.}\ }\textbf {\bibinfo {volume} {18}},\ \bibinfo
  {pages} {13754} (\bibinfo {year} {2016})}\BibitemShut {NoStop}%
\bibitem [{\citenamefont {Artrith}\ \emph {et~al.}(2017)\citenamefont
  {Artrith}, \citenamefont {Urban},\ and\ \citenamefont
  {Ceder}}]{artrith2017efficient}%
  \BibitemOpen
  \bibfield  {author} {\bibinfo {author} {\bibfnamefont {N.}~\bibnamefont
  {Artrith}}, \bibinfo {author} {\bibfnamefont {A.}~\bibnamefont {Urban}}, \
  and\ \bibinfo {author} {\bibfnamefont {G.}~\bibnamefont {Ceder}},\ }\href
  {\doibase 10.1103/PhysRevB.96.014112} {\bibfield  {journal} {\bibinfo
  {journal} {Phys. Rev. B}\ }\textbf {\bibinfo {volume} {96}},\ \bibinfo
  {pages} {014112} (\bibinfo {year} {2017})}\BibitemShut {NoStop}%
\bibitem [{\citenamefont {Bart{\'o}k}\ \emph {et~al.}(2017)\citenamefont
  {Bart{\'o}k}, \citenamefont {De}, \citenamefont {Poelking}, \citenamefont
  {Bernstein}, \citenamefont {Kermode}, \citenamefont {Cs{\'a}nyi},\ and\
  \citenamefont {Ceriotti}}]{Ceriotti2017}%
  \BibitemOpen
  \bibfield  {author} {\bibinfo {author} {\bibfnamefont {A.~P.}\ \bibnamefont
  {Bart{\'o}k}}, \bibinfo {author} {\bibfnamefont {S.}~\bibnamefont {De}},
  \bibinfo {author} {\bibfnamefont {C.}~\bibnamefont {Poelking}}, \bibinfo
  {author} {\bibfnamefont {N.}~\bibnamefont {Bernstein}}, \bibinfo {author}
  {\bibfnamefont {J.~R.}\ \bibnamefont {Kermode}}, \bibinfo {author}
  {\bibfnamefont {G.}~\bibnamefont {Cs{\'a}nyi}}, \ and\ \bibinfo {author}
  {\bibfnamefont {M.}~\bibnamefont {Ceriotti}},\ }\href {\doibase
  10.1126/sciadv.1701816} {\bibfield  {journal} {\bibinfo  {journal} {Sci.
  Adv.}\ }\textbf {\bibinfo {volume} {3}},\ \bibinfo {pages} {e1701816}
  (\bibinfo {year} {2017})}\BibitemShut {NoStop}%
\bibitem [{\citenamefont {Glielmo}\ \emph {et~al.}(2017)\citenamefont
  {Glielmo}, \citenamefont {Sollich},\ and\ \citenamefont
  {De~Vita}}]{Glielmo2017}%
  \BibitemOpen
  \bibfield  {author} {\bibinfo {author} {\bibfnamefont {A.}~\bibnamefont
  {Glielmo}}, \bibinfo {author} {\bibfnamefont {P.}~\bibnamefont {Sollich}}, \
  and\ \bibinfo {author} {\bibfnamefont {A.}~\bibnamefont {De~Vita}},\ }\href
  {\doibase 10.1103/PhysRevB.95.214302} {\bibfield  {journal} {\bibinfo
  {journal} {Phys. Rev. B}\ }\textbf {\bibinfo {volume} {95}},\ \bibinfo
  {pages} {214302} (\bibinfo {year} {2017})}\BibitemShut {NoStop}%
\bibitem [{\citenamefont {Yao}\ \emph {et~al.}(2017)\citenamefont {Yao},
  \citenamefont {Herr},\ and\ \citenamefont {Parkhill}}]{yao2017many}%
  \BibitemOpen
  \bibfield  {author} {\bibinfo {author} {\bibfnamefont {K.}~\bibnamefont
  {Yao}}, \bibinfo {author} {\bibfnamefont {J.~E.}\ \bibnamefont {Herr}}, \
  and\ \bibinfo {author} {\bibfnamefont {J.}~\bibnamefont {Parkhill}},\ }\href
  {\doibase 10.1063/1.4973380} {\bibfield  {journal} {\bibinfo  {journal} {J.
  Chem. Phys.}\ }\textbf {\bibinfo {volume} {146}},\ \bibinfo {pages} {014106}
  (\bibinfo {year} {2017})}\BibitemShut {NoStop}%
\bibitem [{\citenamefont {Faber}\ \emph {et~al.}(2017)\citenamefont {Faber},
  \citenamefont {Hutchison}, \citenamefont {Huang}, \citenamefont {Gilmer},
  \citenamefont {Schoenholz}, \citenamefont {Dahl}, \citenamefont {Vinyals},
  \citenamefont {Kearnes}, \citenamefont {Riley},\ and\ \citenamefont {von
  Lilienfeld}}]{faber2017prediction}%
  \BibitemOpen
  \bibfield  {author} {\bibinfo {author} {\bibfnamefont {F.~A.}\ \bibnamefont
  {Faber}}, \bibinfo {author} {\bibfnamefont {L.}~\bibnamefont {Hutchison}},
  \bibinfo {author} {\bibfnamefont {B.}~\bibnamefont {Huang}}, \bibinfo
  {author} {\bibfnamefont {J.}~\bibnamefont {Gilmer}}, \bibinfo {author}
  {\bibfnamefont {S.~S.}\ \bibnamefont {Schoenholz}}, \bibinfo {author}
  {\bibfnamefont {G.~E.}\ \bibnamefont {Dahl}}, \bibinfo {author}
  {\bibfnamefont {O.}~\bibnamefont {Vinyals}}, \bibinfo {author} {\bibfnamefont
  {S.}~\bibnamefont {Kearnes}}, \bibinfo {author} {\bibfnamefont {P.~F.}\
  \bibnamefont {Riley}}, \ and\ \bibinfo {author} {\bibfnamefont {O.~A.}\
  \bibnamefont {von Lilienfeld}},\ }\href {\doibase 10.1021/acs.jctc.7b00577}
  {\bibfield  {journal} {\bibinfo  {journal} {J. Chem. Theory Comput.}\
  }\textbf {\bibinfo {volume} {13}},\ \bibinfo {pages} {5255} (\bibinfo {year}
  {2017})}\BibitemShut {NoStop}%
\bibitem [{\citenamefont {Eickenberg}\ \emph {et~al.}(2018)\citenamefont
  {Eickenberg}, \citenamefont {Exarchakis}, \citenamefont {Hirn}, \citenamefont
  {Mallat},\ and\ \citenamefont {Thiry}}]{eickenberg2018solid}%
  \BibitemOpen
  \bibfield  {author} {\bibinfo {author} {\bibfnamefont {M.}~\bibnamefont
  {Eickenberg}}, \bibinfo {author} {\bibfnamefont {G.}~\bibnamefont
  {Exarchakis}}, \bibinfo {author} {\bibfnamefont {M.}~\bibnamefont {Hirn}},
  \bibinfo {author} {\bibfnamefont {S.}~\bibnamefont {Mallat}}, \ and\ \bibinfo
  {author} {\bibfnamefont {L.}~\bibnamefont {Thiry}},\ }\href {\doibase
  10.1063/1.5023798} {\bibfield  {journal} {\bibinfo  {journal} {J. Chem.
  Phys.}\ }\textbf {\bibinfo {volume} {148}},\ \bibinfo {pages} {241732}
  (\bibinfo {year} {2018})}\BibitemShut {NoStop}%
\bibitem [{\citenamefont {Glielmo}\ \emph {et~al.}(2018)\citenamefont
  {Glielmo}, \citenamefont {Zeni},\ and\ \citenamefont
  {De~Vita}}]{glielmo2018efficient}%
  \BibitemOpen
  \bibfield  {author} {\bibinfo {author} {\bibfnamefont {A.}~\bibnamefont
  {Glielmo}}, \bibinfo {author} {\bibfnamefont {C.}~\bibnamefont {Zeni}}, \
  and\ \bibinfo {author} {\bibfnamefont {A.}~\bibnamefont {De~Vita}},\ }\href
  {\doibase 10.1103/PhysRevB.97.184307} {\bibfield  {journal} {\bibinfo
  {journal} {Phys. Rev. B}\ }\textbf {\bibinfo {volume} {97}},\ \bibinfo
  {pages} {184307} (\bibinfo {year} {2018})}\BibitemShut {NoStop}%
\bibitem [{\citenamefont {Grisafi}\ \emph {et~al.}(2018)\citenamefont
  {Grisafi}, \citenamefont {Wilkins}, \citenamefont {Cs\'anyi},\ and\
  \citenamefont {Ceriotti}}]{Grisafi2018}%
  \BibitemOpen
  \bibfield  {author} {\bibinfo {author} {\bibfnamefont {A.}~\bibnamefont
  {Grisafi}}, \bibinfo {author} {\bibfnamefont {D.~M.}\ \bibnamefont
  {Wilkins}}, \bibinfo {author} {\bibfnamefont {G.}~\bibnamefont {Cs\'anyi}}, \
  and\ \bibinfo {author} {\bibfnamefont {M.}~\bibnamefont {Ceriotti}},\ }\href
  {\doibase 10.1103/PhysRevLett.120.036002} {\bibfield  {journal} {\bibinfo
  {journal} {Phys. Rev. Lett.}\ }\textbf {\bibinfo {volume} {120}},\ \bibinfo
  {pages} {036002} (\bibinfo {year} {2018})}\BibitemShut {NoStop}%
\bibitem [{\citenamefont {Tang}\ \emph {et~al.}(2018)\citenamefont {Tang},
  \citenamefont {Zhang},\ and\ \citenamefont
  {Karniadakis}}]{tang2018atomistic}%
  \BibitemOpen
  \bibfield  {author} {\bibinfo {author} {\bibfnamefont {Y.-H.}\ \bibnamefont
  {Tang}}, \bibinfo {author} {\bibfnamefont {D.}~\bibnamefont {Zhang}}, \ and\
  \bibinfo {author} {\bibfnamefont {G.~E.}\ \bibnamefont {Karniadakis}},\
  }\href {\doibase 10.1063/1.5008630} {\bibfield  {journal} {\bibinfo
  {journal} {J. Chem. Phys.}\ }\textbf {\bibinfo {volume} {148}},\ \bibinfo
  {pages} {034101} (\bibinfo {year} {2018})}\BibitemShut {NoStop}%
\bibitem [{\citenamefont {Pronobis}\ \emph {et~al.}(2018)\citenamefont
  {Pronobis}, \citenamefont {Tkatchenko},\ and\ \citenamefont
  {M{\"u}ller}}]{pronobis2018many}%
  \BibitemOpen
  \bibfield  {author} {\bibinfo {author} {\bibfnamefont {W.}~\bibnamefont
  {Pronobis}}, \bibinfo {author} {\bibfnamefont {A.}~\bibnamefont
  {Tkatchenko}}, \ and\ \bibinfo {author} {\bibfnamefont {K.-R.}\ \bibnamefont
  {M{\"u}ller}},\ }\href {\doibase 10.1021/acs.jctc.8b00110} {\bibfield
  {journal} {\bibinfo  {journal} {J. Chem. Theory Comput.}\ }\textbf {\bibinfo
  {volume} {14}},\ \bibinfo {pages} {2991} (\bibinfo {year}
  {2018})}\BibitemShut {NoStop}%
\bibitem [{\citenamefont {Faber}\ \emph {et~al.}(2018)\citenamefont {Faber},
  \citenamefont {Christensen}, \citenamefont {Huang},\ and\ \citenamefont {von
  Lilienfeld}}]{FCHL2018}%
  \BibitemOpen
  \bibfield  {author} {\bibinfo {author} {\bibfnamefont {F.~A.}\ \bibnamefont
  {Faber}}, \bibinfo {author} {\bibfnamefont {A.~S.}\ \bibnamefont
  {Christensen}}, \bibinfo {author} {\bibfnamefont {B.}~\bibnamefont {Huang}},
  \ and\ \bibinfo {author} {\bibfnamefont {O.~A.}\ \bibnamefont {von
  Lilienfeld}},\ }\href {\doibase 10.1063/1.5020710} {\bibfield  {journal}
  {\bibinfo  {journal} {J. Chem. Phys.}\ }\textbf {\bibinfo {volume} {148}},\
  \bibinfo {pages} {241717} (\bibinfo {year} {2018})}\BibitemShut {NoStop}%
\bibitem [{\citenamefont {Behler}\ and\ \citenamefont
  {Parrinello}(2007)}]{Behler2007}%
  \BibitemOpen
  \bibfield  {author} {\bibinfo {author} {\bibfnamefont {J.}~\bibnamefont
  {Behler}}\ and\ \bibinfo {author} {\bibfnamefont {M.}~\bibnamefont
  {Parrinello}},\ }\href {\doibase 10.1103/PhysRevLett.98.146401} {\bibfield
  {journal} {\bibinfo  {journal} {Phys. Rev. Lett.}\ }\textbf {\bibinfo
  {volume} {98}},\ \bibinfo {pages} {146401} (\bibinfo {year}
  {2007})}\BibitemShut {NoStop}%
\bibitem [{\citenamefont {Jose}\ \emph {et~al.}(2012)\citenamefont {Jose},
  \citenamefont {Artrith},\ and\ \citenamefont {Behler}}]{Behler2012}%
  \BibitemOpen
  \bibfield  {author} {\bibinfo {author} {\bibfnamefont {K.~V.~J.}\
  \bibnamefont {Jose}}, \bibinfo {author} {\bibfnamefont {N.}~\bibnamefont
  {Artrith}}, \ and\ \bibinfo {author} {\bibfnamefont {J.}~\bibnamefont
  {Behler}},\ }\href {\doibase 10.1063/1.4712397} {\bibfield  {journal}
  {\bibinfo  {journal} {J. Chem. Phys.}\ }\textbf {\bibinfo {volume} {136}},\
  \bibinfo {pages} {194111} (\bibinfo {year} {2012})}\BibitemShut {NoStop}%
\bibitem [{\citenamefont {Behler}(2016)}]{Behler2016}%
  \BibitemOpen
  \bibfield  {author} {\bibinfo {author} {\bibfnamefont {J.}~\bibnamefont
  {Behler}},\ }\href {\doibase 10.1063/1.4966192} {\bibfield  {journal}
  {\bibinfo  {journal} {J. Chem. Phys.}\ }\textbf {\bibinfo {volume} {145}},\
  \bibinfo {pages} {170901} (\bibinfo {year} {2016})}\BibitemShut {NoStop}%
\bibitem [{\citenamefont {Gastegger}\ \emph {et~al.}(2017)\citenamefont
  {Gastegger}, \citenamefont {Behler},\ and\ \citenamefont
  {Marquetand}}]{Gastegger2017}%
  \BibitemOpen
  \bibfield  {author} {\bibinfo {author} {\bibfnamefont {M.}~\bibnamefont
  {Gastegger}}, \bibinfo {author} {\bibfnamefont {J.}~\bibnamefont {Behler}}, \
  and\ \bibinfo {author} {\bibfnamefont {P.}~\bibnamefont {Marquetand}},\
  }\href {\doibase 10.1039/C7SC02267K} {\bibfield  {journal} {\bibinfo
  {journal} {Chem. Sci.}\ }\textbf {\bibinfo {volume} {8}},\ \bibinfo {pages}
  {6924} (\bibinfo {year} {2017})}\BibitemShut {NoStop}%
\bibitem [{\citenamefont {Sch{\"u}tt}\ \emph {et~al.}(2017)\citenamefont
  {Sch{\"u}tt}, \citenamefont {Arbabzadah}, \citenamefont {Chmiela},
  \citenamefont {M{\"u}ller},\ and\ \citenamefont {Tkatchenko}}]{dtnn}%
  \BibitemOpen
  \bibfield  {author} {\bibinfo {author} {\bibfnamefont {K.~T.}\ \bibnamefont
  {Sch{\"u}tt}}, \bibinfo {author} {\bibfnamefont {F.}~\bibnamefont
  {Arbabzadah}}, \bibinfo {author} {\bibfnamefont {S.}~\bibnamefont {Chmiela}},
  \bibinfo {author} {\bibfnamefont {K.~R.}\ \bibnamefont {M{\"u}ller}}, \ and\
  \bibinfo {author} {\bibfnamefont {A.}~\bibnamefont {Tkatchenko}},\ }\href
  {http://dx.doi.org/10.1038/ncomms13890} {\bibfield  {journal} {\bibinfo
  {journal} {Nat. Commun.}\ }\textbf {\bibinfo {volume} {8}},\ \bibinfo {pages}
  {13890} (\bibinfo {year} {2017})}\BibitemShut {NoStop}%
\bibitem [{\citenamefont {Sch\"utt}\ \emph {et~al.}(2018)\citenamefont
  {Sch\"utt}, \citenamefont {Sauceda}, \citenamefont {Kindermans},
  \citenamefont {Tkatchenko},\ and\ \citenamefont {M\"uller}}]{SchNet2018}%
  \BibitemOpen
  \bibfield  {author} {\bibinfo {author} {\bibfnamefont {K.~T.}\ \bibnamefont
  {Sch\"utt}}, \bibinfo {author} {\bibfnamefont {H.~E.}\ \bibnamefont
  {Sauceda}}, \bibinfo {author} {\bibfnamefont {P.-J.}\ \bibnamefont
  {Kindermans}}, \bibinfo {author} {\bibfnamefont {A.}~\bibnamefont
  {Tkatchenko}}, \ and\ \bibinfo {author} {\bibfnamefont {K.-R.}\ \bibnamefont
  {M\"uller}},\ }\href {\doibase 10.1063/1.5019779} {\bibfield  {journal}
  {\bibinfo  {journal} {J. Chem. Phys.}\ }\textbf {\bibinfo {volume} {148}},\
  \bibinfo {pages} {241722} (\bibinfo {year} {2018})}\BibitemShut {NoStop}%
\bibitem [{\citenamefont {Sch\"{u}tt}\ \emph {et~al.}(2017)\citenamefont
  {Sch\"{u}tt}, \citenamefont {Kindermans}, \citenamefont {Sauceda},
  \citenamefont {Chmiela}, \citenamefont {Tkatchenko},\ and\ \citenamefont
  {M\"{u}ller}}]{SchNetNIPS2017}%
  \BibitemOpen
  \bibfield  {author} {\bibinfo {author} {\bibfnamefont {K.~T.}\ \bibnamefont
  {Sch\"{u}tt}}, \bibinfo {author} {\bibfnamefont {P.-J.}\ \bibnamefont
  {Kindermans}}, \bibinfo {author} {\bibfnamefont {H.~E.}\ \bibnamefont
  {Sauceda}}, \bibinfo {author} {\bibfnamefont {S.}~\bibnamefont {Chmiela}},
  \bibinfo {author} {\bibfnamefont {A.}~\bibnamefont {Tkatchenko}}, \ and\
  \bibinfo {author} {\bibfnamefont {K.-R.}\ \bibnamefont {M\"{u}ller}},\ }in\
  \href
  {http://papers.nips.cc/paper/6700-schnet-a-continuous-filter-convolutional-neural-network-for-modeling-quantum-interactions.pdf}
  {\emph {\bibinfo {booktitle} {Advances in Neural Information Processing
  Systems 30}}}\ (\bibinfo  {publisher} {Curran Associates, Inc.},\ \bibinfo
  {year} {2017})\ pp.\ \bibinfo {pages} {991--1001}\BibitemShut {NoStop}%
\bibitem [{\citenamefont {Ryczko}\ \emph {et~al.}(2018)\citenamefont {Ryczko},
  \citenamefont {Mills}, \citenamefont {Luchak}, \citenamefont {Homenick},\
  and\ \citenamefont {Tamblyn}}]{ryczko2018convolutional}%
  \BibitemOpen
  \bibfield  {author} {\bibinfo {author} {\bibfnamefont {K.}~\bibnamefont
  {Ryczko}}, \bibinfo {author} {\bibfnamefont {K.}~\bibnamefont {Mills}},
  \bibinfo {author} {\bibfnamefont {I.}~\bibnamefont {Luchak}}, \bibinfo
  {author} {\bibfnamefont {C.}~\bibnamefont {Homenick}}, \ and\ \bibinfo
  {author} {\bibfnamefont {I.}~\bibnamefont {Tamblyn}},\ }\href {\doibase
  https://doi.org/10.1016/j.commatsci.2018.03.005} {\bibfield  {journal}
  {\bibinfo  {journal} {Comput. Mater. Sci.}\ }\textbf {\bibinfo {volume}
  {149}},\ \bibinfo {pages} {134} (\bibinfo {year} {2018})}\BibitemShut
  {NoStop}%
\bibitem [{\citenamefont {Zhang}\ \emph {et~al.}(2018)\citenamefont {Zhang},
  \citenamefont {Han}, \citenamefont {Wang}, \citenamefont {Car},\ and\
  \citenamefont {Weinan}}]{zhang2018deep}%
  \BibitemOpen
  \bibfield  {author} {\bibinfo {author} {\bibfnamefont {L.}~\bibnamefont
  {Zhang}}, \bibinfo {author} {\bibfnamefont {J.}~\bibnamefont {Han}}, \bibinfo
  {author} {\bibfnamefont {H.}~\bibnamefont {Wang}}, \bibinfo {author}
  {\bibfnamefont {R.}~\bibnamefont {Car}}, \ and\ \bibinfo {author}
  {\bibfnamefont {E.}~\bibnamefont {Weinan}},\ }\href {\doibase
  10.1103/PhysRevLett.120.143001} {\bibfield  {journal} {\bibinfo  {journal}
  {Phys. Rev. Lett.}\ }\textbf {\bibinfo {volume} {120}},\ \bibinfo {pages}
  {143001} (\bibinfo {year} {2018})}\BibitemShut {NoStop}%
\bibitem [{\citenamefont {Li}\ \emph {et~al.}(2015)\citenamefont {Li},
  \citenamefont {Kermode},\ and\ \citenamefont {De~Vita}}]{DeVita2015}%
  \BibitemOpen
  \bibfield  {author} {\bibinfo {author} {\bibfnamefont {Z.}~\bibnamefont
  {Li}}, \bibinfo {author} {\bibfnamefont {J.~R.}\ \bibnamefont {Kermode}}, \
  and\ \bibinfo {author} {\bibfnamefont {A.}~\bibnamefont {De~Vita}},\ }\href
  {\doibase 10.1103/PhysRevLett.114.096405} {\bibfield  {journal} {\bibinfo
  {journal} {Phys. Rev. Lett.}\ }\textbf {\bibinfo {volume} {114}},\ \bibinfo
  {pages} {096405} (\bibinfo {year} {2015})}\BibitemShut {NoStop}%
\bibitem [{\citenamefont {Podryabinkin}\ and\ \citenamefont
  {Shapeev}(2017)}]{Shapeev2017}%
  \BibitemOpen
  \bibfield  {author} {\bibinfo {author} {\bibfnamefont {E.~V.}\ \bibnamefont
  {Podryabinkin}}\ and\ \bibinfo {author} {\bibfnamefont {A.~V.}\ \bibnamefont
  {Shapeev}},\ }\href {\doibase
  https://doi.org/10.1016/j.commatsci.2017.08.031} {\bibfield  {journal}
  {\bibinfo  {journal} {Comput. Mater. Sci.}\ }\textbf {\bibinfo {volume}
  {140}},\ \bibinfo {pages} {171 } (\bibinfo {year} {2017})}\BibitemShut
  {NoStop}%
\bibitem [{\citenamefont {Dral}\ \emph {et~al.}(2017)\citenamefont {Dral},
  \citenamefont {Owens}, \citenamefont {Yurchenko},\ and\ \citenamefont
  {Thiel}}]{dral2017structure}%
  \BibitemOpen
  \bibfield  {author} {\bibinfo {author} {\bibfnamefont {P.~O.}\ \bibnamefont
  {Dral}}, \bibinfo {author} {\bibfnamefont {A.}~\bibnamefont {Owens}},
  \bibinfo {author} {\bibfnamefont {S.~N.}\ \bibnamefont {Yurchenko}}, \ and\
  \bibinfo {author} {\bibfnamefont {W.}~\bibnamefont {Thiel}},\ }\href
  {\doibase 10.1063/1.4989536} {\bibfield  {journal} {\bibinfo  {journal} {J.
  Chem. Phys.}\ }\textbf {\bibinfo {volume} {146}},\ \bibinfo {pages} {244108}
  (\bibinfo {year} {2017})}\BibitemShut {NoStop}%
\bibitem [{\citenamefont {Mardt}\ \emph {et~al.}(2018)\citenamefont {Mardt},
  \citenamefont {Pasquali}, \citenamefont {Wu},\ and\ \citenamefont
  {No{\'e}}}]{noe2018}%
  \BibitemOpen
  \bibfield  {author} {\bibinfo {author} {\bibfnamefont {A.}~\bibnamefont
  {Mardt}}, \bibinfo {author} {\bibfnamefont {L.}~\bibnamefont {Pasquali}},
  \bibinfo {author} {\bibfnamefont {H.}~\bibnamefont {Wu}}, \ and\ \bibinfo
  {author} {\bibfnamefont {F.}~\bibnamefont {No{\'e}}},\ }\href {\doibase
  10.1038/s41467-017-02388-1} {\bibfield  {journal} {\bibinfo  {journal} {Nat.
  Commun.}\ }\textbf {\bibinfo {volume} {9}},\ \bibinfo {pages} {5} (\bibinfo
  {year} {2018})}\BibitemShut {NoStop}%
\bibitem [{\citenamefont {No{\'e}}\ and\ \citenamefont {Wu}(2018)}]{noe2018b}%
  \BibitemOpen
  \bibfield  {author} {\bibinfo {author} {\bibfnamefont {F.}~\bibnamefont
  {No{\'e}}}\ and\ \bibinfo {author} {\bibfnamefont {H.}~\bibnamefont {Wu}},\
  }\href@noop {} {\enquote {\bibinfo {title} {Boltzmann generators - sampling
  equilibrium states of many-body systems with deep learning},}\ } (\bibinfo
  {year} {2018}),\ \Eprint {http://arxiv.org/abs/arXiv:1812.01729}
  {arXiv:1812.01729} \BibitemShut {NoStop}%
\bibitem [{\citenamefont {Bart{\'{o}}k}\ \emph {et~al.}(2013)\citenamefont
  {Bart{\'{o}}k}, \citenamefont {Kondor},\ and\ \citenamefont
  {Cs{\'{a}}nyi}}]{Bartok2013}%
  \BibitemOpen
  \bibfield  {author} {\bibinfo {author} {\bibfnamefont {A.~P.}\ \bibnamefont
  {Bart{\'{o}}k}}, \bibinfo {author} {\bibfnamefont {R.}~\bibnamefont
  {Kondor}}, \ and\ \bibinfo {author} {\bibfnamefont {G.}~\bibnamefont
  {Cs{\'{a}}nyi}},\ }\href {\doibase 10.1103/PhysRevB.87.184115} {\bibfield
  {journal} {\bibinfo  {journal} {Phys. Rev. B}\ }\textbf {\bibinfo {volume}
  {87}},\ \bibinfo {pages} {184115} (\bibinfo {year} {2013})}\BibitemShut
  {NoStop}%
\bibitem [{\citenamefont {Montavon}\ \emph {et~al.}(2013)\citenamefont
  {Montavon}, \citenamefont {Rupp}, \citenamefont {Gobre}, \citenamefont
  {Vazquez-Mayagoitia}, \citenamefont {Hansen}, \citenamefont {Tkatchenko},
  \citenamefont {M{\"{u}}ller},\ and\ \citenamefont {von
  Lilienfeld}}]{Montavon2013a}%
  \BibitemOpen
  \bibfield  {author} {\bibinfo {author} {\bibfnamefont {G.}~\bibnamefont
  {Montavon}}, \bibinfo {author} {\bibfnamefont {M.}~\bibnamefont {Rupp}},
  \bibinfo {author} {\bibfnamefont {V.}~\bibnamefont {Gobre}}, \bibinfo
  {author} {\bibfnamefont {A.}~\bibnamefont {Vazquez-Mayagoitia}}, \bibinfo
  {author} {\bibfnamefont {K.}~\bibnamefont {Hansen}}, \bibinfo {author}
  {\bibfnamefont {A.}~\bibnamefont {Tkatchenko}}, \bibinfo {author}
  {\bibfnamefont {K.-R.}\ \bibnamefont {M{\"{u}}ller}}, \ and\ \bibinfo
  {author} {\bibfnamefont {O.~A.}\ \bibnamefont {von Lilienfeld}},\ }\href
  {http://stacks.iop.org/1367-2630/15/i=9/a=095003} {\bibfield  {journal}
  {\bibinfo  {journal} {New J. Phys.}\ }\textbf {\bibinfo {volume} {15}},\
  \bibinfo {pages} {95003} (\bibinfo {year} {2013})}\BibitemShut {NoStop}%
\bibitem [{\citenamefont {Botu}\ and\ \citenamefont
  {Ramprasad}(2015)}]{Ramprasad2015}%
  \BibitemOpen
  \bibfield  {author} {\bibinfo {author} {\bibfnamefont {V.}~\bibnamefont
  {Botu}}\ and\ \bibinfo {author} {\bibfnamefont {R.}~\bibnamefont
  {Ramprasad}},\ }\href {\doibase 10.1103/PhysRevB.92.094306} {\bibfield
  {journal} {\bibinfo  {journal} {Phys. Rev. B}\ }\textbf {\bibinfo {volume}
  {92}},\ \bibinfo {pages} {094306} (\bibinfo {year} {2015})}\BibitemShut
  {NoStop}%
\bibitem [{\citenamefont {Brockherde}\ \emph {et~al.}(2017)\citenamefont
  {Brockherde}, \citenamefont {Vogt}, \citenamefont {Li}, \citenamefont
  {Tuckerman}, \citenamefont {Burke},\ and\ \citenamefont
  {M{\"u}ller}}]{Brockherde2017}%
  \BibitemOpen
  \bibfield  {author} {\bibinfo {author} {\bibfnamefont {F.}~\bibnamefont
  {Brockherde}}, \bibinfo {author} {\bibfnamefont {L.}~\bibnamefont {Vogt}},
  \bibinfo {author} {\bibfnamefont {L.}~\bibnamefont {Li}}, \bibinfo {author}
  {\bibfnamefont {M.~E.}\ \bibnamefont {Tuckerman}}, \bibinfo {author}
  {\bibfnamefont {K.}~\bibnamefont {Burke}}, \ and\ \bibinfo {author}
  {\bibfnamefont {K.-R.}\ \bibnamefont {M{\"u}ller}},\ }\href {\doibase
  10.1038/s41467-017-00839-3} {\bibfield  {journal} {\bibinfo  {journal} {Nat.
  Commun.}\ }\textbf {\bibinfo {volume} {8}},\ \bibinfo {pages} {872} (\bibinfo
  {year} {2017})}\BibitemShut {NoStop}%
\bibitem [{\citenamefont {Huan}\ \emph {et~al.}(2017)\citenamefont {Huan},
  \citenamefont {Batra}, \citenamefont {Chapman}, \citenamefont {Krishnan},
  \citenamefont {Chen},\ and\ \citenamefont {Ramprasad}}]{huan2017universal}%
  \BibitemOpen
  \bibfield  {author} {\bibinfo {author} {\bibfnamefont {T.~D.}\ \bibnamefont
  {Huan}}, \bibinfo {author} {\bibfnamefont {R.}~\bibnamefont {Batra}},
  \bibinfo {author} {\bibfnamefont {J.}~\bibnamefont {Chapman}}, \bibinfo
  {author} {\bibfnamefont {S.}~\bibnamefont {Krishnan}}, \bibinfo {author}
  {\bibfnamefont {L.}~\bibnamefont {Chen}}, \ and\ \bibinfo {author}
  {\bibfnamefont {R.}~\bibnamefont {Ramprasad}},\ }\href {\doibase
  10.1038/s41524-017-0042-y} {\bibfield  {journal} {\bibinfo  {journal} {{NPJ}
  Comput. Mater.}\ }\textbf {\bibinfo {volume} {3}},\ \bibinfo {pages} {37}
  (\bibinfo {year} {2017})}\BibitemShut {NoStop}%
\bibitem [{\citenamefont {Bereau}\ \emph {et~al.}(2018)\citenamefont {Bereau},
  \citenamefont {DiStasio~Jr}, \citenamefont {Tkatchenko},\ and\ \citenamefont
  {Von~Lilienfeld}}]{Tristan2018}%
  \BibitemOpen
  \bibfield  {author} {\bibinfo {author} {\bibfnamefont {T.}~\bibnamefont
  {Bereau}}, \bibinfo {author} {\bibfnamefont {R.~A.}\ \bibnamefont
  {DiStasio~Jr}}, \bibinfo {author} {\bibfnamefont {A.}~\bibnamefont
  {Tkatchenko}}, \ and\ \bibinfo {author} {\bibfnamefont {O.~A.}\ \bibnamefont
  {Von~Lilienfeld}},\ }\href {\doibase 10.1063/1.5009502} {\bibfield  {journal}
  {\bibinfo  {journal} {J. Chem. Phys.}\ }\textbf {\bibinfo {volume} {148}},\
  \bibinfo {pages} {241706} (\bibinfo {year} {2018})}\BibitemShut {NoStop}%
\bibitem [{\citenamefont {Lubbers}\ \emph {et~al.}(2018)\citenamefont
  {Lubbers}, \citenamefont {Smith},\ and\ \citenamefont
  {Barros}}]{lubbers2018hierarchical}%
  \BibitemOpen
  \bibfield  {author} {\bibinfo {author} {\bibfnamefont {N.}~\bibnamefont
  {Lubbers}}, \bibinfo {author} {\bibfnamefont {J.~S.}\ \bibnamefont {Smith}},
  \ and\ \bibinfo {author} {\bibfnamefont {K.}~\bibnamefont {Barros}},\ }\href
  {\doibase 10.1063/1.5011181} {\bibfield  {journal} {\bibinfo  {journal} {J.
  Chem. Phys.}\ }\textbf {\bibinfo {volume} {148}},\ \bibinfo {pages} {241715}
  (\bibinfo {year} {2018})}\BibitemShut {NoStop}%
\bibitem [{\citenamefont {Kanamori}\ \emph {et~al.}(2018)\citenamefont
  {Kanamori}, \citenamefont {Toyoura}, \citenamefont {Honda}, \citenamefont
  {Hattori}, \citenamefont {Seko}, \citenamefont {Karasuyama}, \citenamefont
  {Shitara}, \citenamefont {Shiga}, \citenamefont {Kuwabara},\ and\
  \citenamefont {Takeuchi}}]{kanamori2018exploring}%
  \BibitemOpen
  \bibfield  {author} {\bibinfo {author} {\bibfnamefont {K.}~\bibnamefont
  {Kanamori}}, \bibinfo {author} {\bibfnamefont {K.}~\bibnamefont {Toyoura}},
  \bibinfo {author} {\bibfnamefont {J.}~\bibnamefont {Honda}}, \bibinfo
  {author} {\bibfnamefont {K.}~\bibnamefont {Hattori}}, \bibinfo {author}
  {\bibfnamefont {A.}~\bibnamefont {Seko}}, \bibinfo {author} {\bibfnamefont
  {M.}~\bibnamefont {Karasuyama}}, \bibinfo {author} {\bibfnamefont
  {K.}~\bibnamefont {Shitara}}, \bibinfo {author} {\bibfnamefont
  {M.}~\bibnamefont {Shiga}}, \bibinfo {author} {\bibfnamefont
  {A.}~\bibnamefont {Kuwabara}}, \ and\ \bibinfo {author} {\bibfnamefont
  {I.}~\bibnamefont {Takeuchi}},\ }\href {\doibase 10.1103/PhysRevB.97.125124}
  {\bibfield  {journal} {\bibinfo  {journal} {Phys. Rev. B}\ }\textbf {\bibinfo
  {volume} {97}},\ \bibinfo {pages} {125124} (\bibinfo {year}
  {2018})}\BibitemShut {NoStop}%
\bibitem [{\citenamefont {Hy}\ \emph {et~al.}(2018)\citenamefont {Hy},
  \citenamefont {Trivedi}, \citenamefont {Pan}, \citenamefont {Anderson},\ and\
  \citenamefont {Kondor}}]{hy2018predicting}%
  \BibitemOpen
  \bibfield  {author} {\bibinfo {author} {\bibfnamefont {T.~S.}\ \bibnamefont
  {Hy}}, \bibinfo {author} {\bibfnamefont {S.}~\bibnamefont {Trivedi}},
  \bibinfo {author} {\bibfnamefont {H.}~\bibnamefont {Pan}}, \bibinfo {author}
  {\bibfnamefont {B.~M.}\ \bibnamefont {Anderson}}, \ and\ \bibinfo {author}
  {\bibfnamefont {R.}~\bibnamefont {Kondor}},\ }\href {\doibase
  10.1063/1.5024797} {\bibfield  {journal} {\bibinfo  {journal} {J. Chem.
  Phys.}\ }\textbf {\bibinfo {volume} {148}},\ \bibinfo {pages} {241745}
  (\bibinfo {year} {2018})}\BibitemShut {NoStop}%
\bibitem [{\citenamefont {Smith}\ \emph {et~al.}(2017)\citenamefont {Smith},
  \citenamefont {Isayev},\ and\ \citenamefont {Roitberg}}]{Smith2017}%
  \BibitemOpen
  \bibfield  {author} {\bibinfo {author} {\bibfnamefont {J.~S.}\ \bibnamefont
  {Smith}}, \bibinfo {author} {\bibfnamefont {O.}~\bibnamefont {Isayev}}, \
  and\ \bibinfo {author} {\bibfnamefont {A.~E.}\ \bibnamefont {Roitberg}},\
  }\href {\doibase 10.1039/C6SC05720A} {\bibfield  {journal} {\bibinfo
  {journal} {Chem. Sci.}\ }\textbf {\bibinfo {volume} {8}},\ \bibinfo {pages}
  {3192} (\bibinfo {year} {2017})}\BibitemShut {NoStop}%
\bibitem [{\citenamefont {Wang}\ \emph {et~al.}(2018)\citenamefont {Wang},
  \citenamefont {Wehmeyer}, \citenamefont {No{\'e}},\ and\ \citenamefont
  {Clementi}}]{Clementi2018}%
  \BibitemOpen
  \bibfield  {author} {\bibinfo {author} {\bibfnamefont {J.}~\bibnamefont
  {Wang}}, \bibinfo {author} {\bibfnamefont {C.}~\bibnamefont {Wehmeyer}},
  \bibinfo {author} {\bibfnamefont {F.}~\bibnamefont {No{\'e}}}, \ and\
  \bibinfo {author} {\bibfnamefont {C.}~\bibnamefont {Clementi}},\ }\href@noop
  {} {\enquote {\bibinfo {title} {Machine learning of coarse-grained molecular
  dynamics force fields},}\ } (\bibinfo {year} {2018}),\ \Eprint
  {http://arxiv.org/abs/arXiv:1812.01736} {arXiv:1812.01736} \BibitemShut
  {NoStop}%
\bibitem [{\citenamefont {Winter}\ \emph {et~al.}(2019)\citenamefont {Winter},
  \citenamefont {Montanari}, \citenamefont {No{\'e}},\ and\ \citenamefont
  {Clevert}}]{winter2019}%
  \BibitemOpen
  \bibfield  {author} {\bibinfo {author} {\bibfnamefont {R.}~\bibnamefont
  {Winter}}, \bibinfo {author} {\bibfnamefont {F.}~\bibnamefont {Montanari}},
  \bibinfo {author} {\bibfnamefont {F.}~\bibnamefont {No{\'e}}}, \ and\
  \bibinfo {author} {\bibfnamefont {D.-A.}\ \bibnamefont {Clevert}},\ }\href
  {\doibase 10.1039/C8SC04175J} {\bibfield  {journal} {\bibinfo  {journal}
  {Chem. Sci.}\ } (\bibinfo {year} {2019}),\ 10.1039/C8SC04175J}\BibitemShut
  {NoStop}%
\bibitem [{\citenamefont {Christensen}\ \emph {et~al.}(2018)\citenamefont
  {Christensen}, \citenamefont {Faber},\ and\ \citenamefont {von
  Lilienfeld}}]{ResponseFieldAnatole2018}%
  \BibitemOpen
  \bibfield  {author} {\bibinfo {author} {\bibfnamefont {A.~S.}\ \bibnamefont
  {Christensen}}, \bibinfo {author} {\bibfnamefont {F.~A.}\ \bibnamefont
  {Faber}}, \ and\ \bibinfo {author} {\bibfnamefont {O.~A.}\ \bibnamefont {von
  Lilienfeld}},\ }\href@noop {} {\enquote {\bibinfo {title} {Operators in
  machine learning: Response properties in chemical space},}\ } (\bibinfo
  {year} {2018}),\ \Eprint {http://arxiv.org/abs/arXiv:1807.08811}
  {arXiv:1807.08811} \BibitemShut {NoStop}%
\bibitem [{\citenamefont {Chmiela}\ \emph {et~al.}(2017)\citenamefont
  {Chmiela}, \citenamefont {Tkatchenko}, \citenamefont {Sauceda}, \citenamefont
  {Poltavsky}, \citenamefont {Sch{\"u}tt},\ and\ \citenamefont
  {M{\"u}ller}}]{gdml}%
  \BibitemOpen
  \bibfield  {author} {\bibinfo {author} {\bibfnamefont {S.}~\bibnamefont
  {Chmiela}}, \bibinfo {author} {\bibfnamefont {A.}~\bibnamefont {Tkatchenko}},
  \bibinfo {author} {\bibfnamefont {H.~E.}\ \bibnamefont {Sauceda}}, \bibinfo
  {author} {\bibfnamefont {I.}~\bibnamefont {Poltavsky}}, \bibinfo {author}
  {\bibfnamefont {K.~T.}\ \bibnamefont {Sch{\"u}tt}}, \ and\ \bibinfo {author}
  {\bibfnamefont {K.-R.}\ \bibnamefont {M{\"u}ller}},\ }\href {\doibase
  10.1126/sciadv.1603015} {\bibfield  {journal} {\bibinfo  {journal} {Sci.
  Adv.}\ }\textbf {\bibinfo {volume} {3}},\ \bibinfo {pages} {e1603015}
  (\bibinfo {year} {2017})}\BibitemShut {NoStop}%
\bibitem [{\citenamefont {Chmiela}\ \emph
  {et~al.}(2018{\natexlab{a}})\citenamefont {Chmiela}, \citenamefont {Sauceda},
  \citenamefont {M{\"u}ller},\ and\ \citenamefont {Tkatchenko}}]{sgdml}%
  \BibitemOpen
  \bibfield  {author} {\bibinfo {author} {\bibfnamefont {S.}~\bibnamefont
  {Chmiela}}, \bibinfo {author} {\bibfnamefont {H.~E.}\ \bibnamefont
  {Sauceda}}, \bibinfo {author} {\bibfnamefont {K.-R.}\ \bibnamefont
  {M{\"u}ller}}, \ and\ \bibinfo {author} {\bibfnamefont {A.}~\bibnamefont
  {Tkatchenko}},\ }\href {\doibase 10.1038/s41467-018-06169-2} {\bibfield
  {journal} {\bibinfo  {journal} {Nat. Commun.}\ }\textbf {\bibinfo {volume}
  {9}},\ \bibinfo {pages} {3887} (\bibinfo {year}
  {2018}{\natexlab{a}})}\BibitemShut {NoStop}%
\bibitem [{\citenamefont {Chmiela}\ \emph
  {et~al.}(2018{\natexlab{b}})\citenamefont {Chmiela}, \citenamefont {Sauceda},
  \citenamefont {Poltavsky}, \citenamefont {M{\"u}ller},\ and\ \citenamefont
  {Tkatchenko}}]{sGDMLsoftware2019}%
  \BibitemOpen
  \bibfield  {author} {\bibinfo {author} {\bibfnamefont {S.}~\bibnamefont
  {Chmiela}}, \bibinfo {author} {\bibfnamefont {H.~E.}\ \bibnamefont
  {Sauceda}}, \bibinfo {author} {\bibfnamefont {I.}~\bibnamefont {Poltavsky}},
  \bibinfo {author} {\bibfnamefont {K.-R.}\ \bibnamefont {M{\"u}ller}}, \ and\
  \bibinfo {author} {\bibfnamefont {A.}~\bibnamefont {Tkatchenko}},\
  }\href@noop {} {\enquote {\bibinfo {title} {sgdml: Constructing accurate and
  data efficient molecular force fields using machine learning},}\ } (\bibinfo
  {year} {2018}{\natexlab{b}}),\ \Eprint
  {http://arxiv.org/abs/arXiv:1812.04986} {arXiv:1812.04986} \BibitemShut
  {NoStop}%
\bibitem [{\citenamefont {Yao}\ \emph {et~al.}(2018)\citenamefont {Yao},
  \citenamefont {Herr}, \citenamefont {Toth}, \citenamefont {Mckintyre},\ and\
  \citenamefont {Parkhill}}]{yao2018tensormol}%
  \BibitemOpen
  \bibfield  {author} {\bibinfo {author} {\bibfnamefont {K.}~\bibnamefont
  {Yao}}, \bibinfo {author} {\bibfnamefont {J.~E.}\ \bibnamefont {Herr}},
  \bibinfo {author} {\bibfnamefont {D.~W.}\ \bibnamefont {Toth}}, \bibinfo
  {author} {\bibfnamefont {R.}~\bibnamefont {Mckintyre}}, \ and\ \bibinfo
  {author} {\bibfnamefont {J.}~\bibnamefont {Parkhill}},\ }\href {\doibase
  10.1039/C7SC04934J} {\bibfield  {journal} {\bibinfo  {journal} {Chem. Sci.}\
  }\textbf {\bibinfo {volume} {9}},\ \bibinfo {pages} {2261} (\bibinfo {year}
  {2018})}\BibitemShut {NoStop}%
\bibitem [{\citenamefont {Sch{\"u}tt}\ \emph {et~al.}(2019)\citenamefont
  {Sch{\"u}tt}, \citenamefont {Kessel}, \citenamefont {Gastegger},
  \citenamefont {Nicoli}, \citenamefont {Tkatchenko},\ and\ \citenamefont
  {M{\"u}ller}}]{schnetpack2018}%
  \BibitemOpen
  \bibfield  {author} {\bibinfo {author} {\bibfnamefont {K.~T.}\ \bibnamefont
  {Sch{\"u}tt}}, \bibinfo {author} {\bibfnamefont {P.}~\bibnamefont {Kessel}},
  \bibinfo {author} {\bibfnamefont {M.}~\bibnamefont {Gastegger}}, \bibinfo
  {author} {\bibfnamefont {K.~A.}\ \bibnamefont {Nicoli}}, \bibinfo {author}
  {\bibfnamefont {A.}~\bibnamefont {Tkatchenko}}, \ and\ \bibinfo {author}
  {\bibfnamefont {K.~R.}\ \bibnamefont {M{\"u}ller}},\ }\href {\doibase
  10.1021/acs.jctc.8b00908} {\bibfield  {journal} {\bibinfo  {journal} {J.
  Chem. Theory Comput.}\ }\textbf {\bibinfo {volume} {15}},\ \bibinfo {pages}
  {448} (\bibinfo {year} {2019})}\BibitemShut {NoStop}%
\bibitem [{\citenamefont {Alber}\ \emph {et~al.}(2018)\citenamefont {Alber},
  \citenamefont {Lapuschkin}, \citenamefont {Seegerer}, \citenamefont
  {Hägele}, \citenamefont {Sch{\"u}tt}, \citenamefont {Montavon},
  \citenamefont {Samek}, \citenamefont {M{\"u}ller}, \citenamefont
  {D{\"a}hne},\ and\ \citenamefont {Kindermans}}]{innvestigate}%
  \BibitemOpen
  \bibfield  {author} {\bibinfo {author} {\bibfnamefont {M.}~\bibnamefont
  {Alber}}, \bibinfo {author} {\bibfnamefont {S.}~\bibnamefont {Lapuschkin}},
  \bibinfo {author} {\bibfnamefont {P.}~\bibnamefont {Seegerer}}, \bibinfo
  {author} {\bibfnamefont {M.}~\bibnamefont {Hägele}}, \bibinfo {author}
  {\bibfnamefont {K.~T.}\ \bibnamefont {Sch{\"u}tt}}, \bibinfo {author}
  {\bibfnamefont {G.}~\bibnamefont {Montavon}}, \bibinfo {author}
  {\bibfnamefont {W.}~\bibnamefont {Samek}}, \bibinfo {author} {\bibfnamefont
  {K.-R.}\ \bibnamefont {M{\"u}ller}}, \bibinfo {author} {\bibfnamefont
  {S.}~\bibnamefont {D{\"a}hne}}, \ and\ \bibinfo {author} {\bibfnamefont
  {P.-J.}\ \bibnamefont {Kindermans}},\ }\href@noop {} {\enquote {\bibinfo
  {title} {i{NN}vestigate neural networks!}}\ } (\bibinfo {year} {2018}),\
  \Eprint {http://arxiv.org/abs/arXiv:1808.04260} {arXiv:1808.04260}
  \BibitemShut {NoStop}%
\bibitem [{\citenamefont {Meila}\ \emph {et~al.}(2018)\citenamefont {Meila},
  \citenamefont {Koelle},\ and\ \citenamefont {Zhang}}]{meila2018}%
  \BibitemOpen
  \bibfield  {author} {\bibinfo {author} {\bibfnamefont {M.}~\bibnamefont
  {Meila}}, \bibinfo {author} {\bibfnamefont {S.}~\bibnamefont {Koelle}}, \
  and\ \bibinfo {author} {\bibfnamefont {H.}~\bibnamefont {Zhang}},\
  }\href@noop {} {\enquote {\bibinfo {title} {A regression approach for
  explaining manifold embedding coordinates},}\ } (\bibinfo {year} {2018}),\
  \Eprint {http://arxiv.org/abs/arXiv:1811.11891} {arXiv:1811.11891}
  \BibitemShut {NoStop}%
\bibitem [{\citenamefont {Friedman}\ \emph {et~al.}(2001)\citenamefont
  {Friedman}, \citenamefont {Hastie},\ and\ \citenamefont {Tibshirani}}]{SLT1}%
  \BibitemOpen
  \bibfield  {author} {\bibinfo {author} {\bibfnamefont {J.}~\bibnamefont
  {Friedman}}, \bibinfo {author} {\bibfnamefont {T.}~\bibnamefont {Hastie}}, \
  and\ \bibinfo {author} {\bibfnamefont {R.}~\bibnamefont {Tibshirani}},\
  }\href@noop {} {\emph {\bibinfo {title} {The elements of statistical
  learning}}},\ Vol.~\bibinfo {volume} {1}\ (\bibinfo  {publisher} {Springer
  series in statistics New York, NY, USA:},\ \bibinfo {year}
  {2001})\BibitemShut {NoStop}%
\bibitem [{\citenamefont {Vapnik}(1995)}]{SLT2}%
  \BibitemOpen
  \bibfield  {author} {\bibinfo {author} {\bibfnamefont {V.~N.}\ \bibnamefont
  {Vapnik}},\ }\href@noop {} {\emph {\bibinfo {title} {The Nature of
  Statistical Learning Theory}}}\ (\bibinfo  {publisher} {Springer-Verlag},\
  \bibinfo {address} {Berlin, Heidelberg},\ \bibinfo {year} {1995})\BibitemShut
  {NoStop}%
\bibitem [{\citenamefont {Behler}\ \emph {et~al.}(2007)\citenamefont {Behler},
  \citenamefont {Lorenz},\ and\ \citenamefont {Reuter}}]{Behler2007b}%
  \BibitemOpen
  \bibfield  {author} {\bibinfo {author} {\bibfnamefont {J.}~\bibnamefont
  {Behler}}, \bibinfo {author} {\bibfnamefont {S.}~\bibnamefont {Lorenz}}, \
  and\ \bibinfo {author} {\bibfnamefont {K.}~\bibnamefont {Reuter}},\ }\href
  {\doibase 10.1063/1.2746232} {\bibfield  {journal} {\bibinfo  {journal} {J.
  Chem. Phys.}\ }\textbf {\bibinfo {volume} {127}},\ \bibinfo {pages} {014705}
  (\bibinfo {year} {2007})}\BibitemShut {NoStop}%
\bibitem [{\citenamefont {Behler}(2011{\natexlab{a}})}]{Behler2011}%
  \BibitemOpen
  \bibfield  {author} {\bibinfo {author} {\bibfnamefont {J.}~\bibnamefont
  {Behler}},\ }\href {\doibase http://dx.doi.org/10.1063/1.3553717} {\bibfield
  {journal} {\bibinfo  {journal} {J. Chem. Phys.}\ }\textbf {\bibinfo {volume}
  {134}},\ \bibinfo {pages} {074106} (\bibinfo {year}
  {2011}{\natexlab{a}})}\BibitemShut {NoStop}%
\bibitem [{\citenamefont {Behler}(2011{\natexlab{b}})}]{Behler2011a}%
  \BibitemOpen
  \bibfield  {author} {\bibinfo {author} {\bibfnamefont {J.}~\bibnamefont
  {Behler}},\ }\href {\doibase 10.1039/C1CP21668F} {\bibfield  {journal}
  {\bibinfo  {journal} {Phys. Chem. Chem. Phys.}\ }\textbf {\bibinfo {volume}
  {13}},\ \bibinfo {pages} {17930} (\bibinfo {year}
  {2011}{\natexlab{b}})}\BibitemShut {NoStop}%
\bibitem [{\citenamefont {Solak}\ \emph {et~al.}(2003)\citenamefont {Solak},
  \citenamefont {Murray-smith}, \citenamefont {Leithead}, \citenamefont
  {Leith},\ and\ \citenamefont {Rasmussen}}]{LearnLinNIPS2002}%
  \BibitemOpen
  \bibfield  {author} {\bibinfo {author} {\bibfnamefont {E.}~\bibnamefont
  {Solak}}, \bibinfo {author} {\bibfnamefont {R.}~\bibnamefont {Murray-smith}},
  \bibinfo {author} {\bibfnamefont {W.~E.}\ \bibnamefont {Leithead}}, \bibinfo
  {author} {\bibfnamefont {D.~J.}\ \bibnamefont {Leith}}, \ and\ \bibinfo
  {author} {\bibfnamefont {C.~E.}\ \bibnamefont {Rasmussen}},\ }in\ \href
  {http://papers.nips.cc/paper/2287-derivative-observations-in-gaussian-process-models-of-dynamic-systems.pdf}
  {\emph {\bibinfo {booktitle} {Advances in Neural Information Processing
  Systems 15}}}\ (\bibinfo  {publisher} {MIT Press},\ \bibinfo {year} {2003})\
  pp.\ \bibinfo {pages} {1057--1064}\BibitemShut {NoStop}%
\bibitem [{\citenamefont {Choudhary}\ \emph {et~al.}(2011)\citenamefont
  {Choudhary}, \citenamefont {Kamer},\ and\ \citenamefont
  {Raines}}]{n2pi_Aspirin2011}%
  \BibitemOpen
  \bibfield  {author} {\bibinfo {author} {\bibfnamefont {A.}~\bibnamefont
  {Choudhary}}, \bibinfo {author} {\bibfnamefont {K.~J.}\ \bibnamefont
  {Kamer}}, \ and\ \bibinfo {author} {\bibfnamefont {R.~T.}\ \bibnamefont
  {Raines}},\ }\href {\doibase 10.1021/jo201389d} {\bibfield  {journal}
  {\bibinfo  {journal} {J. Org. Chem.}\ }\textbf {\bibinfo {volume} {76}},\
  \bibinfo {pages} {7933} (\bibinfo {year} {2011})}\BibitemShut {NoStop}%
\bibitem [{\citenamefont {Blanco}\ and\ \citenamefont
  {L{\'o}pez}(2018)}]{Blanco_n-pi2018}%
  \BibitemOpen
  \bibfield  {author} {\bibinfo {author} {\bibfnamefont {S.}~\bibnamefont
  {Blanco}}\ and\ \bibinfo {author} {\bibfnamefont {J.~C.}\ \bibnamefont
  {L{\'o}pez}},\ }\href {\doibase 10.1021/acs.jpclett.8b01719} {\bibfield
  {journal} {\bibinfo  {journal} {J. Phys. Chem. Lett.}\ }\textbf {\bibinfo
  {volume} {9}},\ \bibinfo {pages} {4632} (\bibinfo {year} {2018})}\BibitemShut
  {NoStop}%
\bibitem [{\citenamefont {Perdew}\ \emph {et~al.}(1996)\citenamefont {Perdew},
  \citenamefont {Burke},\ and\ \citenamefont {Ernzerhof}}]{PBE1996}%
  \BibitemOpen
  \bibfield  {author} {\bibinfo {author} {\bibfnamefont {J.~P.}\ \bibnamefont
  {Perdew}}, \bibinfo {author} {\bibfnamefont {K.}~\bibnamefont {Burke}}, \
  and\ \bibinfo {author} {\bibfnamefont {M.}~\bibnamefont {Ernzerhof}},\ }\href
  {\doibase 10.1103/PhysRevLett.77.3865} {\bibfield  {journal} {\bibinfo
  {journal} {Phys. Rev. Lett.}\ }\textbf {\bibinfo {volume} {77}},\ \bibinfo
  {pages} {3865} (\bibinfo {year} {1996})}\BibitemShut {NoStop}%
\bibitem [{\citenamefont {Tkatchenko}\ and\ \citenamefont
  {Scheffler}(2009)}]{TS}%
  \BibitemOpen
  \bibfield  {author} {\bibinfo {author} {\bibfnamefont {A.}~\bibnamefont
  {Tkatchenko}}\ and\ \bibinfo {author} {\bibfnamefont {M.}~\bibnamefont
  {Scheffler}},\ }\href {\doibase 10.1103/PhysRevLett.102.073005} {\bibfield
  {journal} {\bibinfo  {journal} {Phys. Rev. Lett.}\ }\textbf {\bibinfo
  {volume} {102}},\ \bibinfo {pages} {073005} (\bibinfo {year}
  {2009})}\BibitemShut {NoStop}%
\bibitem [{\citenamefont {Blum}\ \emph {et~al.}(2009)\citenamefont {Blum},
  \citenamefont {Gehrke}, \citenamefont {Hanke}, \citenamefont {Havu},
  \citenamefont {Havu}, \citenamefont {Ren}, \citenamefont {Reuter},\ and\
  \citenamefont {Scheffler}}]{FHIaims2009}%
  \BibitemOpen
  \bibfield  {author} {\bibinfo {author} {\bibfnamefont {V.}~\bibnamefont
  {Blum}}, \bibinfo {author} {\bibfnamefont {R.}~\bibnamefont {Gehrke}},
  \bibinfo {author} {\bibfnamefont {F.}~\bibnamefont {Hanke}}, \bibinfo
  {author} {\bibfnamefont {P.}~\bibnamefont {Havu}}, \bibinfo {author}
  {\bibfnamefont {V.}~\bibnamefont {Havu}}, \bibinfo {author} {\bibfnamefont
  {X.}~\bibnamefont {Ren}}, \bibinfo {author} {\bibfnamefont {K.}~\bibnamefont
  {Reuter}}, \ and\ \bibinfo {author} {\bibfnamefont {M.}~\bibnamefont
  {Scheffler}},\ }\href {\doibase https://doi.org/10.1016/j.cpc.2009.06.022}
  {\bibfield  {journal} {\bibinfo  {journal} {Comput. Phys. Commun.}\ }\textbf
  {\bibinfo {volume} {180}},\ \bibinfo {pages} {2175 } (\bibinfo {year}
  {2009})}\BibitemShut {NoStop}%
\bibitem [{\citenamefont {Turney}\ \emph {et~al.}(2012)\citenamefont {Turney},
  \citenamefont {Simmonett}, \citenamefont {Parrish}, \citenamefont
  {Hohenstein}, \citenamefont {Evangelista}, \citenamefont {Fermann},
  \citenamefont {Mintz}, \citenamefont {Burns}, \citenamefont {Wilke},
  \citenamefont {Abrams}, \citenamefont {Russ}, \citenamefont {Leininger},
  \citenamefont {Janssen}, \citenamefont {Seidl}, \citenamefont {Allen},
  \citenamefont {Schaefer}, \citenamefont {King}, \citenamefont {Valeev},
  \citenamefont {Sherrill},\ and\ \citenamefont {Crawford}}]{psi42012}%
  \BibitemOpen
  \bibfield  {author} {\bibinfo {author} {\bibfnamefont {J.~M.}\ \bibnamefont
  {Turney}}, \bibinfo {author} {\bibfnamefont {A.~C.}\ \bibnamefont
  {Simmonett}}, \bibinfo {author} {\bibfnamefont {R.~M.}\ \bibnamefont
  {Parrish}}, \bibinfo {author} {\bibfnamefont {E.~G.}\ \bibnamefont
  {Hohenstein}}, \bibinfo {author} {\bibfnamefont {F.~A.}\ \bibnamefont
  {Evangelista}}, \bibinfo {author} {\bibfnamefont {J.~T.}\ \bibnamefont
  {Fermann}}, \bibinfo {author} {\bibfnamefont {B.~J.}\ \bibnamefont {Mintz}},
  \bibinfo {author} {\bibfnamefont {L.~A.}\ \bibnamefont {Burns}}, \bibinfo
  {author} {\bibfnamefont {J.~J.}\ \bibnamefont {Wilke}}, \bibinfo {author}
  {\bibfnamefont {M.~L.}\ \bibnamefont {Abrams}}, \bibinfo {author}
  {\bibfnamefont {N.~J.}\ \bibnamefont {Russ}}, \bibinfo {author}
  {\bibfnamefont {M.~L.}\ \bibnamefont {Leininger}}, \bibinfo {author}
  {\bibfnamefont {C.~L.}\ \bibnamefont {Janssen}}, \bibinfo {author}
  {\bibfnamefont {E.~T.}\ \bibnamefont {Seidl}}, \bibinfo {author}
  {\bibfnamefont {W.~D.}\ \bibnamefont {Allen}}, \bibinfo {author}
  {\bibfnamefont {H.~F.}\ \bibnamefont {Schaefer}}, \bibinfo {author}
  {\bibfnamefont {R.~A.}\ \bibnamefont {King}}, \bibinfo {author}
  {\bibfnamefont {E.~F.}\ \bibnamefont {Valeev}}, \bibinfo {author}
  {\bibfnamefont {C.~D.}\ \bibnamefont {Sherrill}}, \ and\ \bibinfo {author}
  {\bibfnamefont {T.~D.}\ \bibnamefont {Crawford}},\ }\href {\doibase
  10.1002/wcms.93} {\bibfield  {journal} {\bibinfo  {journal} {WIREs Comput.
  Mol. Sci.}\ }\textbf {\bibinfo {volume} {2}},\ \bibinfo {pages} {556}
  (\bibinfo {year} {2012})}\BibitemShut {NoStop}%
\bibitem [{\citenamefont {Parrish}\ \emph {et~al.}(2017)\citenamefont
  {Parrish}, \citenamefont {Burns}, \citenamefont {Smith}, \citenamefont
  {Simmonett}, \citenamefont {DePrince}, \citenamefont {Hohenstein},
  \citenamefont {Bozkaya}, \citenamefont {Sokolov}, \citenamefont {Di~Remigio},
  \citenamefont {Richard}, \citenamefont {Gonthier}, \citenamefont {James},
  \citenamefont {McAlexander}, \citenamefont {Kumar}, \citenamefont {Saitow},
  \citenamefont {Wang}, \citenamefont {Pritchard}, \citenamefont {Verma},
  \citenamefont {Schaefer}, \citenamefont {Patkowski}, \citenamefont {King},
  \citenamefont {Valeev}, \citenamefont {Evangelista}, \citenamefont {Turney},
  \citenamefont {Crawford},\ and\ \citenamefont {Sherrill}}]{psi42017}%
  \BibitemOpen
  \bibfield  {author} {\bibinfo {author} {\bibfnamefont {R.~M.}\ \bibnamefont
  {Parrish}}, \bibinfo {author} {\bibfnamefont {L.~A.}\ \bibnamefont {Burns}},
  \bibinfo {author} {\bibfnamefont {D.~G.~A.}\ \bibnamefont {Smith}}, \bibinfo
  {author} {\bibfnamefont {A.~C.}\ \bibnamefont {Simmonett}}, \bibinfo {author}
  {\bibfnamefont {A.~E.}\ \bibnamefont {DePrince}}, \bibinfo {author}
  {\bibfnamefont {E.~G.}\ \bibnamefont {Hohenstein}}, \bibinfo {author}
  {\bibfnamefont {U.}~\bibnamefont {Bozkaya}}, \bibinfo {author} {\bibfnamefont
  {A.~Y.}\ \bibnamefont {Sokolov}}, \bibinfo {author} {\bibfnamefont
  {R.}~\bibnamefont {Di~Remigio}}, \bibinfo {author} {\bibfnamefont {R.~M.}\
  \bibnamefont {Richard}}, \bibinfo {author} {\bibfnamefont {J.~F.}\
  \bibnamefont {Gonthier}}, \bibinfo {author} {\bibfnamefont {A.~M.}\
  \bibnamefont {James}}, \bibinfo {author} {\bibfnamefont {H.~R.}\ \bibnamefont
  {McAlexander}}, \bibinfo {author} {\bibfnamefont {A.}~\bibnamefont {Kumar}},
  \bibinfo {author} {\bibfnamefont {M.}~\bibnamefont {Saitow}}, \bibinfo
  {author} {\bibfnamefont {X.}~\bibnamefont {Wang}}, \bibinfo {author}
  {\bibfnamefont {B.~P.}\ \bibnamefont {Pritchard}}, \bibinfo {author}
  {\bibfnamefont {P.}~\bibnamefont {Verma}}, \bibinfo {author} {\bibfnamefont
  {H.~F.}\ \bibnamefont {Schaefer}}, \bibinfo {author} {\bibfnamefont
  {K.}~\bibnamefont {Patkowski}}, \bibinfo {author} {\bibfnamefont {R.~A.}\
  \bibnamefont {King}}, \bibinfo {author} {\bibfnamefont {E.~F.}\ \bibnamefont
  {Valeev}}, \bibinfo {author} {\bibfnamefont {F.~A.}\ \bibnamefont
  {Evangelista}}, \bibinfo {author} {\bibfnamefont {J.~M.}\ \bibnamefont
  {Turney}}, \bibinfo {author} {\bibfnamefont {T.~D.}\ \bibnamefont
  {Crawford}}, \ and\ \bibinfo {author} {\bibfnamefont {C.~D.}\ \bibnamefont
  {Sherrill}},\ }\href {\doibase 10.1021/acs.jctc.7b00174} {\bibfield
  {journal} {\bibinfo  {journal} {J. Chem. Theory Comput.}\ }\textbf {\bibinfo
  {volume} {13}},\ \bibinfo {pages} {3185} (\bibinfo {year}
  {2017})}\BibitemShut {NoStop}%
\bibitem [{\citenamefont {Smith}\ \emph {et~al.}(2018)\citenamefont {Smith},
  \citenamefont {Burns}, \citenamefont {Sirianni}, \citenamefont {Nascimento},
  \citenamefont {Kumar}, \citenamefont {James}, \citenamefont {Schriber},
  \citenamefont {Zhang}, \citenamefont {Zhang}, \citenamefont {Abbott},
  \citenamefont {Berquist}, \citenamefont {Lechner}, \citenamefont {Cunha},
  \citenamefont {Heide}, \citenamefont {Waldrop}, \citenamefont {Takeshita},
  \citenamefont {Alenaizan}, \citenamefont {Neuhauser}, \citenamefont {King},
  \citenamefont {Simmonett}, \citenamefont {Turney}, \citenamefont {Schaefer},
  \citenamefont {Evangelista}, \citenamefont {DePrince}, \citenamefont
  {Crawford}, \citenamefont {Patkowski},\ and\ \citenamefont
  {Sherrill}}]{psi42018}%
  \BibitemOpen
  \bibfield  {author} {\bibinfo {author} {\bibfnamefont {D.~G.~A.}\
  \bibnamefont {Smith}}, \bibinfo {author} {\bibfnamefont {L.~A.}\ \bibnamefont
  {Burns}}, \bibinfo {author} {\bibfnamefont {D.~A.}\ \bibnamefont {Sirianni}},
  \bibinfo {author} {\bibfnamefont {D.~R.}\ \bibnamefont {Nascimento}},
  \bibinfo {author} {\bibfnamefont {A.}~\bibnamefont {Kumar}}, \bibinfo
  {author} {\bibfnamefont {A.~M.}\ \bibnamefont {James}}, \bibinfo {author}
  {\bibfnamefont {J.~B.}\ \bibnamefont {Schriber}}, \bibinfo {author}
  {\bibfnamefont {T.}~\bibnamefont {Zhang}}, \bibinfo {author} {\bibfnamefont
  {B.}~\bibnamefont {Zhang}}, \bibinfo {author} {\bibfnamefont {A.~S.}\
  \bibnamefont {Abbott}}, \bibinfo {author} {\bibfnamefont {E.~J.}\
  \bibnamefont {Berquist}}, \bibinfo {author} {\bibfnamefont {M.~H.}\
  \bibnamefont {Lechner}}, \bibinfo {author} {\bibfnamefont {L.~A.}\
  \bibnamefont {Cunha}}, \bibinfo {author} {\bibfnamefont {A.~G.}\ \bibnamefont
  {Heide}}, \bibinfo {author} {\bibfnamefont {J.~M.}\ \bibnamefont {Waldrop}},
  \bibinfo {author} {\bibfnamefont {T.~Y.}\ \bibnamefont {Takeshita}}, \bibinfo
  {author} {\bibfnamefont {A.}~\bibnamefont {Alenaizan}}, \bibinfo {author}
  {\bibfnamefont {D.}~\bibnamefont {Neuhauser}}, \bibinfo {author}
  {\bibfnamefont {R.~A.}\ \bibnamefont {King}}, \bibinfo {author}
  {\bibfnamefont {A.~C.}\ \bibnamefont {Simmonett}}, \bibinfo {author}
  {\bibfnamefont {J.~M.}\ \bibnamefont {Turney}}, \bibinfo {author}
  {\bibfnamefont {H.~F.}\ \bibnamefont {Schaefer}}, \bibinfo {author}
  {\bibfnamefont {F.~A.}\ \bibnamefont {Evangelista}}, \bibinfo {author}
  {\bibfnamefont {A.~E.}\ \bibnamefont {DePrince}}, \bibinfo {author}
  {\bibfnamefont {T.~D.}\ \bibnamefont {Crawford}}, \bibinfo {author}
  {\bibfnamefont {K.}~\bibnamefont {Patkowski}}, \ and\ \bibinfo {author}
  {\bibfnamefont {C.~D.}\ \bibnamefont {Sherrill}},\ }\href {\doibase
  10.1021/acs.jctc.8b00286} {\bibfield  {journal} {\bibinfo  {journal} {J.
  Chem. Theory Comput.}\ }\textbf {\bibinfo {volume} {14}},\ \bibinfo {pages}
  {3504} (\bibinfo {year} {2018})}\BibitemShut {NoStop}%
\bibitem [{\citenamefont {Hansen}\ \emph
  {et~al.}(2015{\natexlab{b}})\citenamefont {Hansen}, \citenamefont {Biegler},
  \citenamefont {Ramakrishnan}, \citenamefont {Pronobis}, \citenamefont {von
  Lilienfeld}, \citenamefont {M{\"u}ller},\ and\ \citenamefont
  {Tkatchenko}}]{BoB2015}%
  \BibitemOpen
  \bibfield  {author} {\bibinfo {author} {\bibfnamefont {K.}~\bibnamefont
  {Hansen}}, \bibinfo {author} {\bibfnamefont {F.}~\bibnamefont {Biegler}},
  \bibinfo {author} {\bibfnamefont {R.}~\bibnamefont {Ramakrishnan}}, \bibinfo
  {author} {\bibfnamefont {W.}~\bibnamefont {Pronobis}}, \bibinfo {author}
  {\bibfnamefont {O.~A.}\ \bibnamefont {von Lilienfeld}}, \bibinfo {author}
  {\bibfnamefont {K.-R.}\ \bibnamefont {M{\"u}ller}}, \ and\ \bibinfo {author}
  {\bibfnamefont {A.}~\bibnamefont {Tkatchenko}},\ }\href {\doibase
  10.1021/acs.jpclett.5b00831} {\bibfield  {journal} {\bibinfo  {journal} {J.
  Phys. Chem. Lett.}\ }\textbf {\bibinfo {volume} {6}},\ \bibinfo {pages}
  {2326} (\bibinfo {year} {2015}{\natexlab{b}})}\BibitemShut {NoStop}%
\bibitem [{\citenamefont {Ramakrishnan}\ \emph {et~al.}(2015)\citenamefont
  {Ramakrishnan}, \citenamefont {Dral}, \citenamefont {Rupp},\ and\
  \citenamefont {von Lilienfeld}}]{deltaML2015}%
  \BibitemOpen
  \bibfield  {author} {\bibinfo {author} {\bibfnamefont {R.}~\bibnamefont
  {Ramakrishnan}}, \bibinfo {author} {\bibfnamefont {P.~O.}\ \bibnamefont
  {Dral}}, \bibinfo {author} {\bibfnamefont {M.}~\bibnamefont {Rupp}}, \ and\
  \bibinfo {author} {\bibfnamefont {O.~A.}\ \bibnamefont {von Lilienfeld}},\
  }\href {\doibase 10.1021/acs.jctc.5b00099} {\bibfield  {journal} {\bibinfo
  {journal} {J. Chem. Theory Comput.}\ }\textbf {\bibinfo {volume} {11}},\
  \bibinfo {pages} {2087} (\bibinfo {year} {2015})}\BibitemShut {NoStop}%
\bibitem [{\citenamefont {Huang}\ and\ \citenamefont {von
  Lilienfeld}(2017)}]{amons2017}%
  \BibitemOpen
  \bibfield  {author} {\bibinfo {author} {\bibfnamefont {B.}~\bibnamefont
  {Huang}}\ and\ \bibinfo {author} {\bibfnamefont {O.~A.}\ \bibnamefont {von
  Lilienfeld}},\ }\href@noop {} {\enquote {\bibinfo {title} {The "dna" of
  chemistry: Scalable quantum machine learning with "amons"},}\ } (\bibinfo
  {year} {2017}),\ \Eprint {http://arxiv.org/abs/arXiv:1707.04146}
  {arXiv:1707.04146} \BibitemShut {NoStop}%
\bibitem [{\citenamefont {Guillot}(2002)}]{WaterReview2002}%
  \BibitemOpen
  \bibfield  {author} {\bibinfo {author} {\bibfnamefont {B.}~\bibnamefont
  {Guillot}},\ }\href {\doibase https://doi.org/10.1016/S0167-7322(02)00094-6}
  {\bibfield  {journal} {\bibinfo  {journal} {J. Mol. Liq.}\ }\textbf {\bibinfo
  {volume} {101}},\ \bibinfo {pages} {219 } (\bibinfo {year}
  {2002})}\BibitemShut {NoStop}%
\bibitem [{\citenamefont {Oroguchi}\ and\ \citenamefont
  {Nakasako}(2017)}]{AMBER-Hbond2017}%
  \BibitemOpen
  \bibfield  {author} {\bibinfo {author} {\bibfnamefont {T.}~\bibnamefont
  {Oroguchi}}\ and\ \bibinfo {author} {\bibfnamefont {M.}~\bibnamefont
  {Nakasako}},\ }\href {\doibase 10.1038/s41598-017-16203-w} {\bibfield
  {journal} {\bibinfo  {journal} {Sci. Rep.}\ }\textbf {\bibinfo {volume}
  {7}},\ \bibinfo {pages} {15859} (\bibinfo {year} {2017})}\BibitemShut
  {NoStop}%
\bibitem [{\citenamefont {Scheiner}(2017)}]{Scheiner2017}%
  \BibitemOpen
  \bibfield  {author} {\bibinfo {author} {\bibfnamefont {S.}~\bibnamefont
  {Scheiner}},\ }\href {\doibase 10.3390/molecules22091521} {\bibfield
  {journal} {\bibinfo  {journal} {Molecules}\ }\textbf {\bibinfo {volume}
  {22}},\ \bibinfo {pages} {1521} (\bibinfo {year} {2017})}\BibitemShut
  {NoStop}%
\bibitem [{\citenamefont {Kuhn}\ \emph {et~al.}(2010)\citenamefont {Kuhn},
  \citenamefont {Mohr},\ and\ \citenamefont {Stahl}}]{Kuhn2010}%
  \BibitemOpen
  \bibfield  {author} {\bibinfo {author} {\bibfnamefont {B.}~\bibnamefont
  {Kuhn}}, \bibinfo {author} {\bibfnamefont {P.}~\bibnamefont {Mohr}}, \ and\
  \bibinfo {author} {\bibfnamefont {M.}~\bibnamefont {Stahl}},\ }\href
  {\doibase 10.1021/jm100087s} {\bibfield  {journal} {\bibinfo  {journal} {J.
  Med. Chem.}\ }\textbf {\bibinfo {volume} {53}},\ \bibinfo {pages} {2601}
  (\bibinfo {year} {2010})}\BibitemShut {NoStop}%
\bibitem [{\citenamefont {Hobza}(2002)}]{Hobza2002}%
  \BibitemOpen
  \bibfield  {author} {\bibinfo {author} {\bibfnamefont {P.}~\bibnamefont
  {Hobza}},\ }\href {\doibase 10.1002/qua.10313} {\bibfield  {journal}
  {\bibinfo  {journal} {Int. J. Quantum Chem.}\ }\textbf {\bibinfo {volume}
  {90}},\ \bibinfo {pages} {1071} (\bibinfo {year} {2002})}\BibitemShut
  {NoStop}%
\bibitem [{\citenamefont {Karpfen}\ and\ \citenamefont
  {Kryachko}(2009)}]{Karpfen2009}%
  \BibitemOpen
  \bibfield  {author} {\bibinfo {author} {\bibfnamefont {A.}~\bibnamefont
  {Karpfen}}\ and\ \bibinfo {author} {\bibfnamefont {E.~S.}\ \bibnamefont
  {Kryachko}},\ }\href {\doibase 10.1021/jp9005923} {\bibfield  {journal}
  {\bibinfo  {journal} {J. Phys. Chem. A}\ }\textbf {\bibinfo {volume} {113}},\
  \bibinfo {pages} {5217} (\bibinfo {year} {2009})}\BibitemShut {NoStop}%
\bibitem [{\citenamefont {Wang}\ \emph {et~al.}(2017)\citenamefont {Wang},
  \citenamefont {Danovich}, \citenamefont {Shaik},\ and\ \citenamefont
  {Mo}}]{Changwei2017}%
  \BibitemOpen
  \bibfield  {author} {\bibinfo {author} {\bibfnamefont {C.}~\bibnamefont
  {Wang}}, \bibinfo {author} {\bibfnamefont {D.}~\bibnamefont {Danovich}},
  \bibinfo {author} {\bibfnamefont {S.}~\bibnamefont {Shaik}}, \ and\ \bibinfo
  {author} {\bibfnamefont {Y.}~\bibnamefont {Mo}},\ }\href {\doibase
  10.1021/acs.jctc.6b01133} {\bibfield  {journal} {\bibinfo  {journal} {J.
  Chem. Theory Comput.}\ }\textbf {\bibinfo {volume} {13}},\ \bibinfo {pages}
  {1626} (\bibinfo {year} {2017})}\BibitemShut {NoStop}%
\bibitem [{\citenamefont {Goldsby}\ and\ \citenamefont
  {Chang}(2015)}]{ChangBook}%
  \BibitemOpen
  \bibfield  {author} {\bibinfo {author} {\bibfnamefont {K.}~\bibnamefont
  {Goldsby}}\ and\ \bibinfo {author} {\bibfnamefont {R.}~\bibnamefont
  {Chang}},\ }\href@noop {} {\emph {\bibinfo {title} {Chemistry}}}\ (\bibinfo
  {publisher} {McGraw-Hill Higher Education},\ \bibinfo {year}
  {2015})\BibitemShut {NoStop}%
\bibitem [{\citenamefont {Martyniak}\ \emph {et~al.}(2012)\citenamefont
  {Martyniak}, \citenamefont {Majerz},\ and\ \citenamefont
  {Filarowski}}]{AromaOHN2017}%
  \BibitemOpen
  \bibfield  {author} {\bibinfo {author} {\bibfnamefont {A.}~\bibnamefont
  {Martyniak}}, \bibinfo {author} {\bibfnamefont {I.}~\bibnamefont {Majerz}}, \
  and\ \bibinfo {author} {\bibfnamefont {A.}~\bibnamefont {Filarowski}},\
  }\href {\doibase 10.1039/C2RA20846F} {\bibfield  {journal} {\bibinfo
  {journal} {RSC Adv.}\ }\textbf {\bibinfo {volume} {2}},\ \bibinfo {pages}
  {8135} (\bibinfo {year} {2012})}\BibitemShut {NoStop}%
\bibitem [{\citenamefont {Newberry}\ and\ \citenamefont
  {Raines}(2017)}]{n-pi2017}%
  \BibitemOpen
  \bibfield  {author} {\bibinfo {author} {\bibfnamefont {R.~W.}\ \bibnamefont
  {Newberry}}\ and\ \bibinfo {author} {\bibfnamefont {R.~T.}\ \bibnamefont
  {Raines}},\ }\href {\doibase 10.1021/acs.accounts.7b00121} {\bibfield
  {journal} {\bibinfo  {journal} {Acc. Chem. Res.}\ }\textbf {\bibinfo {volume}
  {50}},\ \bibinfo {pages} {1838} (\bibinfo {year} {2017})}\BibitemShut
  {NoStop}%
\bibitem [{\citenamefont {Deepak}\ and\ \citenamefont
  {Sankararamakrishnan}(2016)}]{AceProNMe2016}%
  \BibitemOpen
  \bibfield  {author} {\bibinfo {author} {\bibfnamefont {R.}~\bibnamefont
  {Deepak}}\ and\ \bibinfo {author} {\bibfnamefont {R.}~\bibnamefont
  {Sankararamakrishnan}},\ }\href {\doibase 10.1016/j.bpj.2016.03.034}
  {\bibfield  {journal} {\bibinfo  {journal} {Biophys. J.}\ }\textbf {\bibinfo
  {volume} {110}},\ \bibinfo {pages} {1967 } (\bibinfo {year}
  {2016})}\BibitemShut {NoStop}%
\bibitem [{\citenamefont {Sarkar}\ \emph {et~al.}(2017)\citenamefont {Sarkar},
  \citenamefont {Reddy}, \citenamefont {Mahapatra},\ and\ \citenamefont
  {K\"oppel}}]{Exp03Spectro2017}%
  \BibitemOpen
  \bibfield  {author} {\bibinfo {author} {\bibfnamefont {R.}~\bibnamefont
  {Sarkar}}, \bibinfo {author} {\bibfnamefont {S.~R.}\ \bibnamefont {Reddy}},
  \bibinfo {author} {\bibfnamefont {S.}~\bibnamefont {Mahapatra}}, \ and\
  \bibinfo {author} {\bibfnamefont {H.}~\bibnamefont {K\"oppel}},\ }\href
  {\doibase 10.1016/j.chemphys.2016.09.011} {\bibfield  {journal} {\bibinfo
  {journal} {Chem. Phys.}\ }\textbf {\bibinfo {volume} {482}},\ \bibinfo
  {pages} {39 } (\bibinfo {year} {2017})}\BibitemShut {NoStop}%
\bibitem [{\citenamefont {Gruene}\ \emph {et~al.}(2008)\citenamefont {Gruene},
  \citenamefont {Rayner}, \citenamefont {Redlich}, \citenamefont {van~der
  Meer}, \citenamefont {Lyon}, \citenamefont {Meijer},\ and\ \citenamefont
  {Fielicke}}]{Fielicke2008}%
  \BibitemOpen
  \bibfield  {author} {\bibinfo {author} {\bibfnamefont {P.}~\bibnamefont
  {Gruene}}, \bibinfo {author} {\bibfnamefont {D.~M.}\ \bibnamefont {Rayner}},
  \bibinfo {author} {\bibfnamefont {B.}~\bibnamefont {Redlich}}, \bibinfo
  {author} {\bibfnamefont {A.~F.~G.}\ \bibnamefont {van~der Meer}}, \bibinfo
  {author} {\bibfnamefont {J.~T.}\ \bibnamefont {Lyon}}, \bibinfo {author}
  {\bibfnamefont {G.}~\bibnamefont {Meijer}}, \ and\ \bibinfo {author}
  {\bibfnamefont {A.}~\bibnamefont {Fielicke}},\ }\href {\doibase
  10.1126/science.1161166} {\bibfield  {journal} {\bibinfo  {journal}
  {Science}\ }\textbf {\bibinfo {volume} {321}},\ \bibinfo {pages} {674}
  (\bibinfo {year} {2008})}\BibitemShut {NoStop}%
\bibitem [{\citenamefont {Romanescu}\ \emph {et~al.}(2012)\citenamefont
  {Romanescu}, \citenamefont {Harding}, \citenamefont {Fielicke},\ and\
  \citenamefont {Wang}}]{Fielicke2012}%
  \BibitemOpen
  \bibfield  {author} {\bibinfo {author} {\bibfnamefont {C.}~\bibnamefont
  {Romanescu}}, \bibinfo {author} {\bibfnamefont {D.~J.}\ \bibnamefont
  {Harding}}, \bibinfo {author} {\bibfnamefont {A.}~\bibnamefont {Fielicke}}, \
  and\ \bibinfo {author} {\bibfnamefont {L.-S.}\ \bibnamefont {Wang}},\ }\href
  {\doibase 10.1063/1.4732308} {\bibfield  {journal} {\bibinfo  {journal} {J.
  Chem. Phys.}\ }\textbf {\bibinfo {volume} {137}},\ \bibinfo {pages} {014317}
  (\bibinfo {year} {2012})}\BibitemShut {NoStop}%
\bibitem [{\citenamefont {Balabin}(2010)}]{Exp01Spectro2010}%
  \BibitemOpen
  \bibfield  {author} {\bibinfo {author} {\bibfnamefont {R.~M.}\ \bibnamefont
  {Balabin}},\ }\href {\doibase 10.1039/b924029b} {\bibfield  {journal}
  {\bibinfo  {journal} {Phys. Chem. Chem. Phys.}\ }\textbf {\bibinfo {volume}
  {12}},\ \bibinfo {pages} {5980} (\bibinfo {year} {2010})}\BibitemShut
  {NoStop}%
\bibitem [{\citenamefont {Ruiz-Santoyo}\ \emph {et~al.}(2016)\citenamefont
  {Ruiz-Santoyo}, \citenamefont {Wilke}, \citenamefont {Wilke}, \citenamefont
  {Yi}, \citenamefont {Pratt}, \citenamefont {Schmitt},\ and\ \citenamefont
  {Álvarez Valtierra}}]{Exp01Spectro2016}%
  \BibitemOpen
  \bibfield  {author} {\bibinfo {author} {\bibfnamefont {J.~A.}\ \bibnamefont
  {Ruiz-Santoyo}}, \bibinfo {author} {\bibfnamefont {J.}~\bibnamefont {Wilke}},
  \bibinfo {author} {\bibfnamefont {M.}~\bibnamefont {Wilke}}, \bibinfo
  {author} {\bibfnamefont {J.~T.}\ \bibnamefont {Yi}}, \bibinfo {author}
  {\bibfnamefont {D.~W.}\ \bibnamefont {Pratt}}, \bibinfo {author}
  {\bibfnamefont {M.}~\bibnamefont {Schmitt}}, \ and\ \bibinfo {author}
  {\bibfnamefont {L.}~\bibnamefont {Álvarez Valtierra}},\ }\href {\doibase
  10.1063/1.4939796} {\bibfield  {journal} {\bibinfo  {journal} {J. Chem.
  Phys.}\ }\textbf {\bibinfo {volume} {144}},\ \bibinfo {pages} {044303}
  (\bibinfo {year} {2016})}\BibitemShut {NoStop}%
\bibitem [{\citenamefont {Davies}\ \emph {et~al.}(2017)\citenamefont {Davies},
  \citenamefont {Whalley},\ and\ \citenamefont {Reid}}]{Exp01Spectro2017}%
  \BibitemOpen
  \bibfield  {author} {\bibinfo {author} {\bibfnamefont {J.~A.}\ \bibnamefont
  {Davies}}, \bibinfo {author} {\bibfnamefont {L.~E.}\ \bibnamefont {Whalley}},
  \ and\ \bibinfo {author} {\bibfnamefont {K.~L.}\ \bibnamefont {Reid}},\
  }\href {\doibase 10.1039/C6CP08132K} {\bibfield  {journal} {\bibinfo
  {journal} {Phys. Chem. Chem. Phys.}\ }\textbf {\bibinfo {volume} {19}},\
  \bibinfo {pages} {5051} (\bibinfo {year} {2017})}\BibitemShut {NoStop}%
\bibitem [{\citenamefont {Gmerek}\ \emph {et~al.}(2017)\citenamefont {Gmerek},
  \citenamefont {Stuhlmann}, \citenamefont {Pehlivanovic},\ and\ \citenamefont
  {Schmitt}}]{Exp02Spectro2017}%
  \BibitemOpen
  \bibfield  {author} {\bibinfo {author} {\bibfnamefont {F.}~\bibnamefont
  {Gmerek}}, \bibinfo {author} {\bibfnamefont {B.}~\bibnamefont {Stuhlmann}},
  \bibinfo {author} {\bibfnamefont {E.}~\bibnamefont {Pehlivanovic}}, \ and\
  \bibinfo {author} {\bibfnamefont {M.}~\bibnamefont {Schmitt}},\ }\href
  {\doibase 10.1016/j.molstruc.2017.04.092} {\bibfield  {journal} {\bibinfo
  {journal} {J. Mol. Struct}\ }\textbf {\bibinfo {volume} {1143}},\ \bibinfo
  {pages} {265 } (\bibinfo {year} {2017})}\BibitemShut {NoStop}%
\end{thebibliography}%
%


\end{document}